\documentclass{article}
\usepackage{microtype}
\usepackage{graphicx}
\usepackage{subfigure}
\usepackage{comment}
\usepackage{longtable}
\usepackage{booktabs} 

\usepackage[accepted]{icml2025}
\usepackage{amsmath}
\usepackage{amssymb}
\usepackage{mathtools}
\usepackage{amsthm}
\usepackage{enumitem}
\usepackage{balance}

\usepackage[hidelinks,
            colorlinks=true,
            linkcolor=blue,
            urlcolor=blue,
            citecolor=blue]{hyperref}
\usepackage[capitalize,noabbrev]{cleveref}
\usepackage[hyphens]{url}
\usepackage[hyphenbreaks]{breakurl}

\icmltitlerunning{Stop Treating `AGI' as the North-star Goal of AI Research}

\newcommand{\icmlLeadContribution}{\& Lead contributors}
\newcommand{\icmlTopContribution}{\textsuperscript{\textdagger}Top contributors}
\newcommand{\icmlContribution}{\textsuperscript{\textdaggerdbl}Contributors (alphabetical order)}
\newcommand{\icmlSeniorContribution}{\textsuperscript{\S}Senior contributors}

\icmlsetsymbol{equal}{*}
\icmlsetsymbol{lead}{$\filledstar$}
\icmlsetsymbol{top}{\textdagger}
\icmlsetsymbol{cont}{\textdaggerdbl}
\icmlsetsymbol{senior}{\S}

\begin{document}

\twocolumn[
\icmltitle{Position: Stop Treating `AGI' as the North-star Goal of AI Research}

\begin{icmlauthorlist}
\icmlauthor{Borhane Blili-Hamelin}{equal,ARVA}
\icmlauthor{Christopher Graziul}{equal,uc}
\icmlauthor{Leif Hancox-Li}{top,vijil}
\icmlauthor{Hananel Hazan}{top,tufts}
\icmlauthor{El-Mahdi El-Mhamdi}{cont,ca,ep} 
\icmlauthor{Avijit Ghosh}{cont,hf,uconn}
\icmlauthor{Katherine Heller}{cont,goog}
\icmlauthor{Jacob Metcalf}{cont,ds}
\icmlauthor{Fabricio Murai}{cont,wpi}
\icmlauthor{Eryk Salvaggio}{cont,rit}
\icmlauthor{Andrew Smart}{cont,goog}
\icmlauthor{Todd Snider}{cont,tu}
\icmlauthor{Mariame Tighanimine}{cont,cn}
\icmlauthor{Talia Ringer}{senior,uiuc}
\icmlauthor{Margaret Mitchell}{senior,hf}
\icmlauthor{Shiri Dori-Hacohen}{senior,uconn}
\end{icmlauthorlist}

\icmlaffiliation{ARVA}{AI Risk and Vulnerability Alliance, NY, USA}
\icmlaffiliation{uc}{University of Chicago, IL, USA}
\icmlaffiliation{tufts}{Tufts University, MA, USA}
\icmlaffiliation{vijil}{Vijil, CA, USA}
\icmlaffiliation{goog}{Google, CA, USA}
\icmlaffiliation{wpi}{Worcester Polytechnic Institute, MA, USA}
\icmlaffiliation{ep}{Ecole Polytechnique, France}
\icmlaffiliation{ca}{Calicarpa, Switzerland}
\icmlaffiliation{cn}{Conservatoire national des arts et métiers, Lise‐CNRS, France}
\icmlaffiliation{tu}{Eberhard Karls Universit\"at T\"ubingen, T\"ubingen, Germany}
\icmlaffiliation{uiuc}{University of Illinois Urbana-Champaign, IL, USA}
\icmlaffiliation{hf}{Hugging Face, NY, USA}
\icmlaffiliation{uconn}{University of Connecticut, CT, USA}
\icmlaffiliation{ds}{Data \& Society Research Institute, NY, USA}
\icmlaffiliation{rit}{Rochester Institute of Technology, NY, USA}
\vskip 0.2in
\icmlcorrespondingauthor{Borhane Blili-Hamelin}{borhane@avidml.org}
\icmlcorrespondingauthor{Christopher Graziul}{graziul@uchicago.edu}
\icmlcorrespondingauthor{Talia Ringer}{tringer@illinois.edu}
\icmlcorrespondingauthor{Margaret Mitchell}{meg@huggingface.co}
\icmlcorrespondingauthor{Shiri Dori-Hacohen}{shiridh@uconn.edu}

\icmlkeywords{Scientific Methodology, Artificial General Intelligence}
]
\vskip 0.2in 

\printAffiliationsAndNotice{\icmlEqualContribution
\icmlLeadContribution{}
\icmlTopContribution{}
\icmlContribution{}
\icmlSeniorContribution{}
}
 
\begin{abstract}

The AI research community plays a vital role in shaping the scientific, engineering, and societal goals of AI research. In this position paper, we argue that focusing on the highly contested topic of `artificial general intelligence' (`AGI') undermines our ability to choose effective goals. We identify six key traps---obstacles to productive goal setting---that are aggravated by AGI discourse: Illusion of Consensus, Supercharging Bad Science, Presuming Value-Neutrality, Goal Lottery, Generality Debt, and Normalized Exclusion. To avoid these traps, we argue that the AI research community needs to (1) prioritize \textbf{specificity} in scientific, engineering, and societal goals, (2) center \textbf{pluralism} about multiple worthwhile approaches to multiple valuable goals, and (3) foster innovation through greater \textbf{inclusion} of disciplines and communities. Therefore, the AI research community needs to \textbf{stop treating ``AGI'' as the north-star goal of AI research}.

\end{abstract}

\section{Introduction}

How can we ensure that AI research goals serve scientific, engineering, and societal needs? What constitutes good science in AI research? Who gets to shape AI research goals? What makes a research goal legitimate or worthwhile? In this position paper, we argue that a widespread emphasis on AGI threatens to undermine the ability of researchers to provide well-motivated answers to these questions. 

Recent advances in large language models (LLMs) have sparked interest in ``achieving human-level `intelligence''' as a ``north-star goal'' of the AI field \cite{mccarthy_proposal_1955,morris_position_2024}. This goal is often referred to as ``artificial general intelligence'' (``AGI'') \cite{chollet_openai_2024, tibebu_deepseek_2025}. Yet rather than helping the field converge around shared goals, AGI discourse has mired it in controversies. Researchers diverge on what AGI is and assumptions about goals and risks \cite{summerfield_natural_2023, morris_position_2024, blili-hamelin_unsocial_2024}. Researchers further contest the motivations, incentives, values, and scientific standing of claims about AGI \cite{gebru_tescreal_2024,mitchell_debates_2024,ahmed_field-building_2024,altmeyer_position_2024}. Finally, the building blocks of AGI as a concept---intelligence and generality---are contested in their own right \cite{gould_mismeasure_1981, anderson_situated_2002, hernandez2018paradigms, cave_problem_2020, raji_ai_2021, alexandrova_democratising_2022, blili-hamelin_making_2023, KarenHaoPanel_2023, guest_metatheory_2024, pmlr-v235-paolo24a, mueller2024myth}.

Building on prior work on the ambiguity between exploratory and confirmatory research in ML \cite{herrmann_position_2024}, unscientific performance claims \cite{altmeyer_position_2024}, SOTA-chasing \cite{raji_ai_2021, church_emerging_2022}, homogenization of research approaches \cite{kleinberg_algorithmic2021, fishman_should_2022, bommasani_picking_2022}, the values embedded in ML research \cite{birhane_values_2022}, and more, our account identifies key obstacles to productive goal setting in AI research---\textbf{traps}.\footnote{Our terminology parallels \citet{selbst_fairness_2019} on fairness.} We provide them here as a diagnosis of problems in goal-setting that we believe are normatively worth addressing, but that the AGI narrative makes difficult to overcome.  To avoid these traps, \textbf{we posit that communities should stop treating AGI as the north-star goal\footnotemark{} of AI research}. 

\footnotetext{Sailors who navigate by the astronomical North Star use it to orient their travels toward a desired destination on Earth. With AGI, some researchers are using AGI as a ``guiding star'' to orient their AI research ``travels'' towards. Other researchers, however, are actually hoping and working towards the goal of ``arriving'' at AGI. Our paper argues against both of these approaches.}

An overarching theme in our discussion is the research community's \emph{unique responsibility to help distinguish hype from reality}. The outputs of AI research are deployed as real-world products at a staggering pace, in proliferating contexts, affecting billions of people. This warrants urgent work on trusted, evidence-based answers to questions about the scientific, engineering, and societal merits of AI tools. As argued by the U.N.'s AI Advisory Body, there is ``an overwhelming amount of information\textellipsis~making it difficult to decipher hype from reality. This can fuel confusion, forestall common understanding and advantage major AI companies at the expense of policymakers, civil society and the public'' \cite{united_nations_governing_2024}. Our position paper addresses this theme: Each of the six traps in our account is an obstacle to distinguishing hype from reality. 

A secondary theme in our discussion is the relationship between people and technology. Ultimately, we argue that instead of a single north-star goal, the AI community needs to pursue \emph{multiple specific} scientific, engineering, and societal goals. If building consensus around an alternative unifying goal proves useful, we propose the goal of \emph{supporting and benefiting human beings}. 

In the next section, we examine six `traps' in AI research---obstacles to productive goal setting (\S\ref{sec:traps}). We argue that AGI discourse reinforces and amplifies each problem. Subsequently, we provide three recommendations for avoiding these traps (\S\ref{sec:recommendations}): specificity of goals; pluralism of goals and approaches; and more inclusive goal setting. We conclude by offering a rebuttal against an \emph{alternative view} (\S\ref{sec:alternative}): that AGI should remain the north-star goal of the field. 

Because the contested nature of AGI is a central theme in the present paper, we avoid providing our own definition for the term. Instead, we provide example definitions throughout the present discussion, as well as a table of illustrative definitions in~\cref{sec:definition}.\footnote{Arguably, many of the concerns we raise about AGI apply to other terms used to refer to future forms of AI, such as ``powerful AI" \cite{machines-grace} and ``transformative AI" \cite{gruetzemacher_transformative_2022}. Ultimately, as we revisit in \S\ref{sec:alternative}, our account can be viewed as critically interrogating north-star goals more generally.} 
We detail how AGI currently serves as a north-star goal in~\cref{sec:north-star}.

\section{Traps}
\label{sec:traps}
We examine six key \emph{traps} that hinder the research community's ability to set worthwhile goals. We argue that each is aggravated by AGI narratives. 

The problems we discuss are highly interrelated. For instance, SOTA-chasing, discussed in relationship to the role of misaligned incentives in shaping goal setting (\S\ref{sec:lottery}), also has implications for bad science (\S\ref{sec:pseudoscience}). Similarly, the problem of the lack of consensus about AGI (\S\ref{sec:illusion}) is a theme that recurs throughout. We do not intend the traps to be mutually exclusive. Rather, our goal for each trap is to provide distinct and useful insights for mitigating failure modes in productive goal setting.

\vspace{-0.075in}
\subsection{Illusion of Consensus}
\label{sec:illusion}

\emph{Using shared term(s) in a way that gives a false impression of consensus about goals, despite goals being contested}

The popular use of the term ``AGI'' \cite{venturebeat2023agi,ibm2023agi,reuters2025trumpai} creates a sense of familiarity, giving the illusion that there is a shared understanding on what AGI is, and broad agreement on research goals in AGI development. However, there are vastly different opinions on what the term AGI refers to, what an AGI research agenda looks like, and what the goals in AGI development are.  Left unchecked, this illusion obstructs explicit engagement on what the goals of AI research are and should be.

From popular discourse to research papers to corporate marketing materials, the vast majority of references to AGI fall into this trap when they uncritically cite claims about so-called AGI. For examples of uncritical media claims, see \citet{venturebeat2023agi} and \citet{ibm2023agi}; see \citet{altmeyer_position_2024} for examples of overhyped research.\footnote{Some researchers who advocate for AGI as a goal have avoided the Illusion of Consensus trap (\S\ref{sec:illusion}); e.g., \citet{morris_position_2024} explicitly call for investigating disagreements about goals, predictions, and risks that underpin prominent accounts of AGI.}
\citet{summerfield_natural_2023} summarizes the issue: 
``AI researchers hope to discover how to build AGI. The problem is that nobody really knows exactly what an AGI would look like.'' \citet{mueller2024myth} calls AGI ``a meaningless concept, an emperor with no clothes.'' \citet{blili-hamelin_unsocial_2024} identify multiple types of disagreement among definitions of AGI or human-level AI. The contested nature of AGI as a goal is even more acute in critiques of AGI concepts \citep[e.g.,][]{altmeyer_position_2024,mueller2024myth,van2024reclaiming}.

Beyond AGI, AI research is rife with topics that involve disagreement about goals, values, and concepts. For example, \citet{mulligan_privacy_2016} argue that \emph{privacy} should be understood as an ``essentially contested concept.'' They argue lack of agreement about the meaning and significance of privacy is not merely a matter of confusion---rather, disagreement and contestation are desirable features that enable privacy to adapt to changing technical and social contexts. Similarly, there is now widespread acceptance \emph{fairness} should be understood as a contested topic, not only admitting incompatible mathematical formalizations but also incompatible values, worldviews, and theoretical assumptions \cite{friedler_impossibility_2021,jacobs_measurement_2021}. 

In suggesting this trap, we do not presume that contested topics are inherently problematic. Rather, we argue that when dealing with the important question of the goals of AI research, the significant disagreements that surround AGI should be embraced as signals of conflicting values. 

\vspace{-0.075in}
\subsection{Supercharging Bad Science}
\label{sec:pseudoscience}

\emph{Worsening current problems with bad science in AI due to poorly defined concepts and experimental procedures}

Research that produces reliable empirical knowledge about AI is vital to public interest decisions about AI's potential for societal and environmental benefit and harm. Yet, many experts have noted a pervasive lack of scientific grounding in AI research \cite{hullman_worst, raji_fallacy_2022, sloane_silicon_2022, suchman_uncontroversial_2023, guest_metatheory_2024, narayanan_ai_2024, united_nations_governing_2024, van2024reclaiming, widder_watching_2024}. We argue that vagueness in AGI discourse exacerbates existing problems with the scientific validity of AI research.

\textbf{Problem 1: Underspecification and external validity.} One problem with the pursuit of AGI as a concrete goal is \textbf{underspecification} \cite{JMLR:v23:20-1335}, where \emph{lack of specificity in goals or concepts leads to cascading epistemic problems}, including irrefutability, lack of external validity, flawed experimental design, and flawed evaluation. These common problems in AI research are worsened in the AGI context by the lack of scientifically grounded definitions of AGI (\S\ref{sec:illusion}). 

Underspecification of learning goals also undermines \emph{external validity}---the question of whether a measurement corresponds to the real-world phenomenon it's supposed to capture. A good example is the debate about whether ``language understanding'' benchmarks actually measure language understanding \cite{jacobs_measurement_2021, liao_are_2021}.

External validity is also relevant when researchers equate human faculties with model proxies \cite{hullman_worst}, such as claiming that a model ``capable of linking specific objects with more general visual context'' is evidence of ``imagination'' \cite{fei_towards_2022}. This rhetorical move is enabled by using colloquial terms like ``imagination'' without considering whether it corresponds to the human faculty. 
\citet{altmeyer_position_2024} likewise critiques \citet{gurnee2024languagemodelsrepresentspace}
for inflated claims enabled by the vagueness of the term ``world model''. Underspecified goals trickle down into many areas of experimental design, such as learning pipelines, evaluation metrics, tasks, representations, and methods.

External validity is also undermined when researchers claim to measure concepts from other fields, like intelligence.
The fields of psychology, neuroscience, and cognitive science have studied human intelligence for generations, yet even they lack consensus on what ``intelligence'' is \cite{gopnik_ais_2019,KarenHaoPanel_2023}. Conversely, AI research is no longer concerned with modeling human cognition \cite{guest_metatheory_2024,van2024reclaiming}. Instead, AI developers define ``intelligence'' on their own terms, privileging definitions convenient for benchmarking or selling products (\S\ref{sec:lottery}), while benefiting from historically positive connotations of the term ``intelligence''. 
 
\textbf{Problem 2: Ambiguity between science and engineering.} Another problem with the pursuit of AGI is \emph{confusion between science and engineering} \cite{agre2014toward, hutchinson_evalgaps, altmeyer_position_2024}. As \citet{hullman_worst} point out, a ``typical supervised ML paper'' (e.g., one that reports accuracy metrics on a benchmark) is often just an ``engineering artifact'', a tool attached to performance claims that cannot be refuted because of replication challenges. \citet{altmeyer_position_2024} argue that this ambiguity between science and engineering means rigorous hypothesis testing with ``specific conditions and considering effect sizes'' is often omitted, with results often presented as ``engineering achievements'' without specifying \emph{precisely} what is being tested, relevant hypotheses, and what effect sizes would constitute substantial findings. 

This ambiguity invites experimenter and confirmation biases, since researchers are incentivized to ``pay little or no attention to competing hypotheses or explanations'' or ``[fail] to articulate a sufficiently strong null hypothesis,'' \cite{altmeyer_position_2024}. Confusion between science and engineering also manifests when it is unclear if a study is pursuing scientific goals---of explanation, hypothesis confirmation, etc.---or goals of specific engineering applications---e.g., a proof-of-concept \cite{hutchinson_evalgaps}. This exacerbates questions about external validity: without clear and specific experimental goals, it is easier to provide post-hoc interpretations of experiments that ``support'' a wide variety of goals (\S\ref{sec:lottery}).

\textbf{Problem 3: Ambiguity between confirmatory and exploratory research.} The ambiguity between engineering and scientific methodology is related to another problem: \emph{confusion between confirmatory and exploratory research} \cite{pmlr-v97-bouthillier19a, herrmann_position_2024}. \citet{herrmann_position_2024} state that confirmatory research ``aims to test preexisting hypotheses to confirm or refute existing theories [while] exploratory research is an open-ended approach that aims to gain insight and understanding in a new or unexplored area.'' They go on to argue that ``most current empirical machine learning research is fashioned as confirmatory research while it should rather be considered exploratory'' and that experiments are ``set up to \emph{confirm} the (implicit) hypothesis that the proposed method constitutes an improvement'' (emphasis theirs). By implicitly conflating exploratory analysis with confirmatory research, ``exploratory findings have a slippery way of `transforming' into planned findings as the research process progresses'' \cite{calin2019new}. Using the vague and contested concept of AGI to frame confirmatory claims worsens this problem, as it makes it harder to figure out \emph{what} is being claimed.

\vspace{-0.075in}
\subsection{Presuming Value-Neutrality}
\label{sec:valueladen}

\emph{Framing goals as purely technical or scientific, when they are in fact laden with political, social, or ethical values}

Presuming Value-Neutrality occurs when technical or scientific goals become disconnected from their \textbf{value-laden} assumptions: aspects of AI research that are---and should be---informed by political, social, and ethical considerations. The AI research community has recently begun examining these value-laden assumptions \cite{shilton_values_2018, broussard_artificial_2019, abebe_roles_2020, blodgett_language_2020, costanza-chock_design_2020, denton_bringing_2020, denton2021genealogy, dotan_value-laden_2020, birhane_towards_2021, green_data_2021, scheuerman2021datasets, viljoen_relational_2021, birhane_values_2022, bommasani_evaluation_2023, fishman_should_2022, hutchinson_evalgaps, mathur_disordering_2022, blili-hamelin_making_2023, blili-hamelin_unsocial_2024, pmlr-v235-zhao24a}. 

When efforts to define AGI and related concepts do not explicitly examine the societal goals and values embedded in their definitions, they fall into the Presuming Value-Neutrality trap. Examples include proposals for ``universal intelligence'' \cite{legg_universal_2007, HERNANDEZORALLO201450}.  

The pursuit of value-neutral approaches echoes debates about psychometric views of human intelligence. Intelligence, like ``health,'' and ``well-being,'' inherently carries normative assumptions about which behaviors or abilities are desirable \citep{anderson_situated_2002,alexandrova_democratising_2022}. Researchers fall into the Presuming Value-Neutrality trap by sidestepping these value-laden dimensions \cite{anderson_situated_2002, cave_problem_2020, blili-hamelin_making_2023}. \citet{warne_spearmans_2019} exemplify this by advocating for purely statistical definitions of intelligence, precisely because cultural definitions vary. 

Value-laden assumptions within concepts like AGI drive legitimate disagreement about their meaning, reflecting divergent societal goals \cite{blili-hamelin_unsocial_2024}. This makes consensus on AGI challenging, as it requires alignment on political, social, and ethical priorities. Similar disagreements affect related concepts like AI \cite{cave_problem_2020, blili-hamelin_making_2023}, ``human-level AI'', ``superintelligence'', and ``strong AI'', reinforcing the Illusion of Consensus trap (\S\ref{sec:illusion}). 

\vspace{-0.075in}
\subsection{Goal Lottery}
\label{sec:lottery}

\emph{Adopting goals which are not adequately justified by scientific, engineering, or social merit, but instead on the basis of incentives, circumstances, or luck}

Researchers have studied the role of socioeconomic factors, trends, and circumstantial factors in shaping AI research. For instance, \citet{hooker_hardware_2021} has argued that a form of hardware lottery---the greater availability of hardware with strengths in parallel processing---was key to the resurgence of deep learning in the 2010s.\footnote{On similar lottery or path dependence effects, see \citet{liebowitz_path_1995, peacock_path_2009, dehghani_benchmark_2021, fishman_should_2022, rossbach_innocent_2023, bauer_mirror_2024,hooker_diminishing_2024}.} Similarly, researchers have examined the role of incentives, socioeconomic factors, and hype cycles in AI research \cite{raji_fallacy_2022,delgado_participatory_2023,sartori_minding_2023,widder_dislocated_2023,gebru_tescreal_2024,hicks_chatgpt_2024,narayanan_ai_2024,wang_against_2024}. With this trap, we focus on cases where lotteries (luck) or incentives drive the adoption of unjustified goals---goals that are inadequately supported by scientific, engineering, or societal merit. 

Consider AGI definitions centered on economic value, like OpenAI's emphasis on ``outperform[ing] humans at most economically valuable work'' \cite{openai_2018}. The primacy of economic value for setting AI research goals is contentious from both engineering and societal perspectives. Such definitions create misalignment between incentives and justifications by reducing complex societal, engineering, and scientific considerations to purely economic metrics.

Another example is benchmark SOTA-chasing---pursuing top scores on popular benchmarks  \cite{bender_dangers_2021, raji_ai_2021,church_emerging_2022, hullman_worst}. Despite strong professional incentives encouraging this practice, it lacks scientific, engineering, and societal justification. Benchmarks poorly reflect model performance in real application contexts because of problems like data leakage, overfitting to benchmarks, and data heterogeneity \cite{el2021collaborative, el2023impossible, hanneke2022no, balloccu-etal-2024-leak, xu2024benchmarkingbenchmarkleakagelarge, zhang2024careful}. In short, the measurement method lacks \emph{external validity}. Yet the practice persists due to reputational and financial rewards, demonstrating misalignment between incentivized goals and their
actual merits. 

The dynamics of goal lotteries are also visible in the story of the multi-decade neglect of deep learning architectures. In this case, a research agenda was sidelined for reasons that eventually proved to be misguided from an engineering, scientific, or societal perspective \citep[e.g., due to ``gatekeeping'' effects against less popular research agendas; see][]{SilerEtAl2014}. Meanwhile, the AI industry went all in on the expert systems ``bubble'' \cite{haigh-bust}. \emph{Reductions in diversity within} contemporary AI research can be a sign that similar mistakes are at play (\S\ref{sec:exclusion}). Some recent initiatives to counter homogenization \cite{chollet_arc_technical_report_2024} rely on operationalizing AGI through benchmarks.\footnote{\citeauthor{chollet_so_2024} proposes that ``We will have AGI when creating [benchmarks `that are easy for humans, yet impossible for AI'] becomes outright impossible'' (\citeyear{chollet_so_2024}).} In practice, they end up as yet another benchmark: incentivizing SOTA-chasing, supercharged by intense media and marketing attention \cite{jones_how_2025}. For this reason, we remain somewhat skeptical of whether approaches like ARC \cite{chollet_measure_2019} outweigh the negative consequences of news-cycle-accelerated SOTA-chasing.

\vspace{-0.075in}
\subsection{Generality Debt}  
\label{sec:generality}

\emph{Relying on the generality or flexibility of tools to postpone crucial engineering, scientific, or societal decisions}

AGI definitions differ on how much ``generality'' is desirable \cite{blili-hamelin_unsocial_2024}. This indicates a lack of clarity and consensus about the goals of AI research, forming a trap that (a) encourages suboptimal science/engineering practices (related to points made in \ref{sec:pseudoscience}); (b) suppresses important social/ethical questions about which research directions are worth pursuing. We term this trap ``Generality Debt'' to parallel technical debt \cite{sculley2014machine}: it delays the work that needs to be done as part of AI research which, if left undone, takes more work to address in the future.

This trap includes the appeal to many different notions of generality at play in machine learning: (1) variety of tasks \cite{hernandez2018paradigms}; (2) capability to be trained for ``any task'' vs.\ ability to perform many predefined tasks \cite{hernandez2018paradigms}; (3) whether the task or data distribution the model is being evaluated on is ``seen'' or ``unseen'' (i.e., available, or not, to the model during its training phase) \cite{altmeyer_position_2024}; (4) variety of data in model input/output, such as structured vs unstructured, modality, etc.; (5) whether the performance of the model reflects ``performance considered `surprising' to humans'' \cite{altmeyer_position_2024}; (6) variety of goals; (7) ability to ``accept a general language for the problem statement'' \cite{newell1965search}; and (8) having a ``general'' internal representation \cite{newell1965search, mccarthy1981some}.

As \citet{pmlr-v235-paolo24a} note, despite the multiple possible meanings of ``generality'', most papers do not define generality even if it is central to their argument. Without formal definition, assessing or improving generalization becomes challenging. Assuming that ``generalization'' is desirable while acknowledging its poor definition is misguided. We should first define specific, measurable properties before arguing that they are desirable.

\textbf{Without proper definition, the value of generality remains unclear.} Different types of generality support different future visions, raising unexplored questions about their relative importance. Further, vague definitions of generality lead to bad science and engineering (\S\ref{sec:pseudoscience}). For example, \citet{altmeyer_position_2024} note how the pursuit of generality has led to vague task specifications. In parallel, \citet{gebru_tescreal_2024} argue that some conceptions of AGI contravene good engineering practices: it is hard to test the functionality of systems under ``standard operating conditions'' if the system is advertised as a ``universal algorithm for learning and acting in any environment.''

Aiming to achieve certain forms of generality could also mean making a tradeoff with ecological validity, as argued by \citet{saxon2024benchmarksmicroscopesmodelmetrology}. They argue that, in practice, ``holistic'' benchmarks tend to be a collection of disparate specific benchmark tasks, meaning that they have task-level construct validity. However, these tasks do not always match with \emph{user-relevant capabilities}. Methodological challenges to achieving such capabilities may or may not be overcome, in time, given innovative solutions, but the pursuit of AGI assumes such challenges are surmountable. Methodological issues often inspire novel solutions, but we cannot assume a solution will be found, nor can those pursuing AGI. Further, strategies for pursuing AGI may introduce or reveal new challenges to achieving user-relevant capabilities.

Finally, the vagueness around ``generality'' is also an ethical concern, as it sidesteps normative questions about \emph{which types of generality merit pursuit} and obscures implicit decisions about how to prioritize different research directions. 

\vspace{-0.075in}
\subsection{Normalized Exclusion}
\label{sec:exclusion}

\emph{Excluding communities and experts from shaping goals}

The negative consequences of exclusion in AI have been extensively discussed, both in terms of how it affects product quality and model performance \citep[e.g.,][]{obermeyer_dissecting_2019}, and also how it harms people \citep[e.g.,][]{buolamwini_gender_2018,shelby_sociotechnical_2023,whitney_real_2024}. We argue that AGI discourse aggravates problems of exclusion. 

\textbf{Problem 1: Excluding communities.} Many communities are left out of meaningful participation in shaping the goals of AI research \cite{delgado_participatory_2023}. Excluding communities causes serious harm, especially to minoritized communities \cite{pierregetting_2021}; it also undermines the utility of end products, reduces model performance, oversimplifies technical challenges \cite{kierans2024quantifyingmisalignmentagentssociotechnical}, and can impede innovation \citep[e.g.,][]{burt_structural_2004}. For instance, the infamous case of facial recognition engines---such as those used by Google, Apple, and Meta---mistaking Black people for gorillas \cite{bbcGoogleApologises} is still occurring more than 8 years after the problem was first identified \cite{racismandtechnologyRacistTechnology,nytimesGooglesPhoto}, with downstream impacts on surveillance and law enforcement \cite{jones2020law,pour2023police}. Similarly, selective forms of inclusion in data annotation raise ethical and practical concerns about downstream effects \cite{wang_whose_2022,bertelsen_data_2024}. Other examples of exclusion or inclusion of communities impacting performance are plentiful \citep[see][]{buolamwini_gender_2018, young_beyond_2019, raji_saving_2020, andrews_reanimation_2024, bergman_stela_2024, salavati_reducing_2024, weidinger_star_2024}. Excluding communities from meaningful feedback also undermines societal goals, such as fostering collective legitimacy through accountability to impacted communities \cite{schulz_addressing_2002,mikesell_ethical_2013, costanza-chock_design_2020,birhane_power_2022,young_participation_2024}. 

The recent prominence of AGI discourse intensifies the existing problem of community exclusion in AI research \cite{frank2017machines,cnn2024muskai}. Many proponents of AGI envision a future where AI systems perform an extraordinary range of tasks for countless communities. However, research and design processes fall short of the inclusiveness demanded by this ambitious vision. For example, December 2024 reporting suggests that OpenAI and Microsoft ``signed an agreement last year stating OpenAI has only achieved AGI when it develops AI systems that can generate at least \$100 billion in profits'' \cite{OpenAIStructure,zeff2024microsoft}, a stark departure from OpenAI's public definition of AGI \cite{openai_2018}. As several authors argue, economic value is not the only type of desirable value \cite{agrawal_large_2022,dulka_use_2022,harrigian_characterization_2023,morris_position_2024,pierson_using_2025}. The economic definition helps guide the engineering decisions of OpenAI. However, it is questionable whether an emphasis on profits will lead to the most beneficial or useful end products or to meaningful consensus about goals, especially for minoritized groups. 

\textbf{Problem 2: Excluding disciplines.} From application domains (e.g., medicine \cite{obermeyer_dissecting_2019}, finance \cite{cao_finance2022}, cybersecurity \cite{salem_advancing_2024}, learning \cite{leong_testing_2024}) to the practices involved in building AI---data annotation, qualitative and quantitative methods, domain expertise, computer science, and many more \cite{wang_whose_2022,bertelsen_data_2024,widder_epistemic_2024}---AI research crosses disciplinary boundaries. The cross-disciplinary challenges of AI research mirror those of other disciplines \cite{stokols2003evaluating,stirling2014disciplinary,amoo2020breaking,vestal2020interdisciplinarity,royal2024science}. 

One major challenge is disciplinary silos, where knowledge is inadequately shared across disciplines \cite{stokols2003evaluating,stirling2014disciplinary,ballantyne_epistemic_2019,amoo2020breaking,dipaolo_whats_2022,royal2024science}. For instance, lack of knowledge sharing could be partly responsible for low attention to the distinction between explanatory and exploratory research in ML, discussed in \S\ref{sec:pseudoscience} \cite{herrmann_position_2024}. 

Another challenge is epistemic hierarchies---where the expertise of some disciplines is explicitly or implicitly devalued \cite{knorr_cetina_epistemic_1999,knorr_cetina_culture_2007,simonton_psychologys_2004,fourcade_superiority_2015, graziul_does_2023}. This can manifest as expert groups being limited to narrow input rather than contributing to broader research design decisions \cite{bertelsen_data_2024}. 

AI researchers' focus on applying their work to other domains creates another major challenge. Insufficient domain knowledge might affect the functionality of AI tools---whether they operate as advertised \cite{raji_fallacy_2022}. For instance, AI tools are deployed to make predictions about future individual-level outcomes, from pre-trial risk prediction and predictive policing to automated employment decisions \cite{wang_against_2024}. Yet inadequate evidence of effectiveness often fails to prevent predictive tools from being built, marketed, and deployed \cite{doucette_impact_2021,cameron_us_2023,connealy_staggered_2024}.

The problem of disciplinary silos becomes particularly acute in AGI-oriented research due to two factors: its claims to be creating cognates or replacements of human intelligence, and its claims to expertise in many disciplines. In the former case, claims are often made while ignoring debates in cognitive science and psychology about the nature of intelligence \cite{summerfield_natural_2023,guest_metatheory_2024,mitchell_debates_2024,van2024reclaiming}. In the latter case, achieving AGI is often framed in terms of being able to ``replace'' domain experts in various domains---which are then often taken up uncritically by the media without input from domain experts themselves \citep[e.g.,][]{henshall_when_2024}.

\textbf{Problem 3: Resource disparities.} In recent years, we have witnessed an unprecedented growth in computational resources required for model training, with requirements doubling approximately every few months \cite{Sevilla_2022}. This trend compounds existing resource disparities, as state-of-the-art AI research often relies on access to computational resources accessible to very few researchers \cite{yu_ai_2023,laforge_dangers_2024}. The financial cost of these resources excludes a wide range of actors from contributing to AI research, as even top research universities have a fraction of the computational resources that many corporations use to advance AI research. Efforts are underway to address this resource disparity by supporting access to large-scale computational resources maintained by government entities (e.g., NAIRR in the United States). Yet, resource parity is an aspirational goal in response to widespread recognition that AI researchers in industry enjoy a \textit{de facto} advantage in setting the goals of AI research due to their access to industrial scale computational resources. This structural advantage is reinforced by the use of pre-print archives to publicize AI research without peer review \cite{devlin2019bert,rombach2022highresolutionimagesynthesislatent,bubeck2023sparksartificialgeneralintelligence,openai2024gpt4technicalreport}, a strategy which legitimizes this work as scientific in nature without applying traditional standards for scientific integrity \cite{tenopir_trustworthiness_2016,lin_how_2020, soderberg_credibility_2020, rastogi_arxiv_2022,kwon_use_2025}.

While resource disparities exist for all forms of AI research, they are particularly stark when AGI is taken as a north-star goal for the discipline, due to the orientation of current AGI efforts towards sheer computational scale and the concentration of such efforts in large tech companies.\footnote{Large-scale efforts also have detrimental impacts on climate, reinforcing resource disparities \cite{bucknall_current_2022,kaack2022aligning,luccioni2024light}.} Such concentration of power makes it even more important that those efforts include, rather than exclude, relevant communities and experts. That is, AGI discourse accelerates the existing trend in AI of discounting domain expertise and lived experiences in favor of models that are allegedly experts in everything. 

\vspace{-0.1in}
\section{Recommendations}\label{sec:recommendations}
We have argued that AGI discourse hinders setting well-motivated scientific and engineering goals in AI development, while being destructive to the development of AI that has social merit. We now provide three recommendations for avoiding these traps.

\textbf{Recommendation 1: Goal Specificity.}
\emph{The AI community must prioritize \textbf{highly specific} language when discussing the scientific, engineering, and societal goals of AI.}

More specific definitions of tangible scientific, engineering, and societal goals promote a shared understanding of these goals, and thus the capacity to evaluate whether these goals are well-motivated. Without such specificity, researchers, practitioners, and others can develop divergent understandings of a goal and how it should be achieved. This divergence enables conceptual arbitrage on the part of AI researchers and practitioners who seek to advance their own goals, since these actors ultimately determine the details of model development and implementation. People external to AI development may then be left assuming that a system achieves a specific goal when, in fact, it does not.

Specificity can maintain sufficient flexibility for exploratory research by engaging in best practices around developing a research question. Consider the research goal ``How can a mixture of experts (MoE) strategy improve performance of Whisper large-v3 in challenging speech domains?'' which could cover various domains of speech or MoE strategies. A reformulated, more specific goal would be: “How can a MoE strategy, where experts are a series of models fine-tuned on short ($<$2s), medium (2--20s), and long utterances ($>$20s), help improve performance of Whisper large-v3 in a speech domain dominated by short utterances but also containing relatively long utterances?” Without additional specification, answering the first question provides little guarantee that a solution would address the specific features of the second question, at least not without substantial effort to understand how such a general solution may be applied to this specific case/domain. This example is illustrative, though based on real features of a speech domain where accuracy is essential (i.e., police radio communications, see~\citealt{srivastava_speech_2024,venkit_race_2024}). 

Goal Specificity addresses the Illusion of Consensus trap (\S\ref{sec:illusion}), promoting conceptual clarity as an essential part of goal-setting. Clarity also helps to avoid the Goal Lottery trap (\S\ref{sec:lottery}) by making goal selection explicit. It similarly addresses the Presuming Value-Neutrality Trap (\S\ref{sec:valueladen}) by explicitly surfacing values tied to specific goals, and directly reduces Generality Debt (\S\ref{sec:generality}) . Finally, goal specificity addresses the underspecification issues highlighted in the Supercharging Bad Science Trap (\S\ref{sec:pseudoscience}).

\textbf{Recommendation 2: Pluralism of goals and approaches.}
\emph{Rather than a single general north-star goal (or small set of goals), the AI community should articulate \textbf{many} worthwhile scientific, engineering, and societal goals---and many possible paths to fulfilling them.}

Reaching meaningful scientific and societal consensus on the goals of a field as broad-ranging as AI is challenging. When consensus may not be viable or desirable, we recommend pluralism: allowing multiple viable conceptions of the goals of AI research. Pluralism is healthy in a society composed of individuals and institutions with divergent values. By default, the research community should be pluralistic about goals and paths to achieving them, aiming for heterogeneity instead of homogeneity \cite{sorensen_value_2024,sorensen_roadmap_2024}.\footnote{We're not rejecting consensus on unifying, general north-star goals as a matter of principle. In some circumstances, like coordinating collective action in response to the climate crisis, strong consensus may become crucial. But the research community should not begin from the assumption that such consensus is necessary, or that consensus is optimal from a scientific or societal perspective.}

Researchers who study the dynamics of knowledge production and problem-solving in groups have found pluralism to be beneficial \cite{hong_groups_2004,muldoon_diversity_2013}, including unique benefits ascribable to egalitarian group dynamics \cite{xu_flat_2022}. 

Pluralism in AI research can take several forms, each with distinct implications for how we approach complex problems. For example:

\textbf{Methodological pluralism} implies that different ways of approaching a problem help achieve better solutions, often an effective strategy for complex problem-solving in general \cite{midgley_methodological_2000,veit_model_2020, zhu_paradigm_2022} .

\textbf{Value pluralism}, as applied to alignment research, implies a direct connection between technical advancement and accommodation of plural values \cite{sorensen_value_2024,sorensen_roadmap_2024}.

\textbf{Algorithmic pluralism} addresses the “patterned inequality” associated with algorithmic monoculture and implies a “plurality of paths to different outcomes” must be supported to avoid reproducing existing social inequalities (e.g., embedded in data) \cite{jain_algorithmic_2024} .


Each form of pluralism translates into concrete research practices, such as choice of method(s), setting of pluralistic goals, and testing algorithms to ensure known harms (i.e., algorithmic discrimination) are addressed. 

To name just one example, algorithmic decision-making in hiring processes is unlikely to benefit from AGI as much as from targeted solutions to that particular setting, which account for the domain-specific nature of most positions, the different ways hiring managers evaluate candidates, and existing evidence of algorithmic discrimination in hiring. In practice, pluralism can manifest in how resources are distributed among different research approaches. For example, rather than investing most computational resources in pursuit of AGI, they could be more evenly distributed among diverse goals and approaches within AI.

Pluralism addresses the Illusion of Consensus trap (\S\ref{sec:illusion}) by acknowledging the lack of consensus, and using diversity of perspectives as a tool for scientific and social progress; the Goal Lottery trap (\S\ref{sec:lottery}) by reducing the chances of arbitrarily or prematurely excluding some goals from consideration; and the Exclusion trap (\S\ref{sec:exclusion}) by encouraging a plurality of goals and approaches.

\textbf{Recommendation 3: Greater Inclusion in Goal Setting.}
\emph{Greater inclusion of communities and disciplines in shaping the goals of AI research is beneficial to innovation.}

Inclusion supports innovation \cite{burt_structural_2004,hewlett_how_2013,zhang_relationship_2021,xu_flat_2022}. Identifying worthwhile goals, related use cases, and potential unintended consequences depends on engaging diverse viewpoints. Within AI research, these viewpoints must include those of end users, experts from other fields, those affected by research outcomes, and data annotators. Excluding any of these groups impoverishes the potential of AI to achieve worthwhile goals since it would discount the perspectives that define these goals as worthwhile. Including them enriches AI research. 

Cross-pollination of ideas between disciplines leads to more impactful research \citep{dorihacohen2021fairness,shi_surprising_2023}. Such impact requires that we abandon silos of (tacit) knowledge \citep[e.g., epistemic cultures:][]{knorr_cetina_epistemic_1999,knorr_cetina_culture_2007} and prioritize epistemic hierarchies that value non-computational research \citep{simonton_psychologys_2004,fourcade_superiority_2015}. While technical complexity can make participation by non-experts challenging \cite{pierregetting_2021}, working through these challenges can surface issues experts have not anticipated \citep{cooper_systematic_2022} and enable practical scientific contributions by integrating the insights and experiences of non-experts, or experts in other fields, into system design decisions \citep{delgado_participatory_2023,salavati_reducing_2024}.

As a topic, AGI often involves imagining AI technologies that impact the lives of everyone. Exclusion thus causes socially significant disagreements regarding the goals, processes, and actors who shape AI research and deployment. These disagreements are often overlooked or ignored by those with the power to shape the field. Inclusion is necessary to ensure that decisions about technology are sufficiently justified to institutions, communities, and individuals \cite{anderson_epistemology_2006,binns_algorithmic_2018,alexandrova_democratising_2022,birhane_power_2022,lazar_power_2022,ovadya_reimagining_2023}.

This recommendation addresses the Normalized Exclusion trap (\S\ref{sec:exclusion}) by treating inclusion as essential to innovation. Moreover, it addresses the Illusion of Consensus and the Presuming Value-Neutrality traps (\S\ref{sec:illusion}, \S\ref{sec:valueladen}) by acknowledging socially significant disagreements about the value-laden goals of AI research and development and using those disagreements productively.

\vspace{-0.1in}
\section{Alternative Views}
\label{sec:alternative}
We have argued that AGI is a poor choice as a north-star to guide AI research. We conclude by championing our position against a strong alternative view: that the traps we have identified can be addressed through a modified pursuit of AGI. We argue that improved approaches to AGI would not go far enough. 

\vspace{-0.075in}
\subsection{Counterargument}
``\emph{AGI is a good north-star goal; to avoid the above traps, improved definitions and accounts of AGI are needed}.''

Thoughtful attempts to address shortcomings in accounts of AGI indeed exist \citep[e.g.,][]{adams2012mapping,chollet_measure_2019,summerfield_natural_2023,morris_position_2024}. If prior accounts of AGI are counterproductive or flawed, why not pursue new accounts that address those flaws? 

As an example of an alternative view,~\citet{morris_position_2024} arguably mitigate the Illusion of Consensus trap by disentangling the disagreements about goals, predictions, and risks that plague other accounts of AGI. Moreover, their proposal for a practical strategy analogous to Levels of Driving Automation standards \cite{sae_international_j3016_202104_2021} could be viewed as mitigating the Presuming Value-Neutrality trap. Setting society-wide standards could be done in a way that explicitly centers \emph{specific} risks that the standards address \cite{morris_position_2024}. In this way, the values being favored manifest in the risks that are centered by the standards. Such works showcase a potential response to traps. In arguing that AGI discourse aggravates multiple standing problems, we cannot rule out the possibility of efforts that mitigate these same problems while retaining AGI as a goal. Moreover, given that lack of agreement about how to define AGI is likely to persist, as \citet{summerfield_natural_2023}, \citet{morris_position_2024}, and many others believe, it would be especially implausible for us to presume to have a complete enough view of the landscape of possible conceptions of AGI to draw definitive conclusions. 

\vspace{-0.075in}
\subsection{Rebuttal}
Why favor our position against this alternative? 

\textbf{Reason 1: Conflict with our recommendations.} Although the definition of AGI is highly contested, a frequent motivation for embracing AGI as a north-star goal is the desire for a single, large-scale, unifying vision for the field \cite{summerfield_natural_2023}. This is somewhat in tension with our recommendation of goal pluralism, which we argue is valuable for AI research. ``AGI''  also brings with it a notion of ``general'' that can be discordant with our recommendation of specificity.\footnote{Note that these are not \emph{all things considered} (or definitive) reasons to abandon the alternative view. Rather, readers who are convinced by our arguments in favor of specificity and pluralism have, to some extent, reasons to be wary of the project of improving AGI. These are so-called \textit{pro-tanto} (i.e., ``to that extent'') reasons, which can be overridden by other considerations \cite{sep-reasons-just-vs-expl}.} As such, there are reasons to be wary of AGI-focused alternatives.

\textbf{Reason 2: Distinguishing hype from reality.} Another reason to favor our position is the AI research community's responsibility to help distinguish hype from reality. We believe that our community must provide trusted, evidence-based answers to increasingly complex questions about AI technologies, their goals, and their impacts. In the current moment, the hyped terminology of AGI undermines this responsibility. 

No matter how well or poorly defined, AGI has acquired a cultural significance that exacerbates the challenge of distinguishing hype from reality. ``Intelligence'' and ``generality'' hold the promise of being beneficial for countless needs and contexts (\S\ref{sec:valueladen}, \S\ref{sec:generality}). No matter how cautious the research community attempts to be, the cultural associations of these terms risk stoking the flames of unscientific thinking about AI. This enables various parties to loosely project utopian or dystopian characteristics onto AGI in ways that support their calls for more power and resources. 

\textbf{Reason 3: Benefiting humans as the goal of technology.} If the AI community nevertheless wants an overarching goal to strive towards, the goal should be the support and benefit of human beings. The goals of technology are shaped by \emph{people}. Evidence-based approaches to examining whether technology effectively meets the needs of people---be they ``users'', ``consumers'', ``patients'', ``scholars'', or a myriad of business and social monikers---are well-established. In a quest to achieve AGI, communities often lose sight of the needs of people as a goal, in favor of focusing on just the technology.

There is another, more ambitious reason to work towards consensus on supporting and benefiting human beings as a goal. We have noted the role of socially significant disagreements about the goals of technology in our third recommendation of inclusion. Processes ensuring that technology benefits humans have the potential to provide \emph{collectively legitimate} responses to such disagreements. This could occur through processes that embrace democratic ideals: such as universal inclusion in interrogation, deliberation, and dissent about the ``common good'' and ``public interest'', while enacting strong accountability to participants as ``rights-holders'' \cite{anderson_epistemology_2006, putnam_reconsideration_2011, binns_algorithmic_2018, gabriel_artificial_2020, birhane_power_2022, lazar_ai_2023, ovadya_reimagining_2023, blili-hamelin_unsocial_2024}. Aiming for collective legitimacy amounts to requiring politically and socially effective forms of consensus. 

In sum, we urge communities to \textbf{stop treating ``AGI'' as the north-star goal of AI research}.

\vspace{-0.1in}
\section*{Acknowledgments}
Borhane Blili-Hamelin was funded in part through a Magic Grant from the Brown Institute for Media Innovation. Christopher Graziul was supported by the National Institute of Minority Health and Health Disparities of the National Institutes of Health under award number R01MD015064. This material is based upon work supported in part by the NSF Program on Fairness in AI in Collaboration with Amazon under Award IIS-2147305. Any opinions, findings, and conclusions or recommendations expressed in this material are those of the author(s) and do not necessarily reflect the views of the National Science Foundation, National Institutes of Health, the Brown Institute for Media Innovation, or Amazon.

\sloppy
\bibliographystyle{icml2025}
\balance
\vspace{-0.1in}
\bibliography{references}

\begin{thebibliography}{224}
\providecommand{\natexlab}[1]{#1}
\providecommand{\url}[1]{\texttt{#1}}
\expandafter\ifx\csname urlstyle\endcsname\relax
  \providecommand{\doi}[1]{doi: #1}\else
  \providecommand{\doi}{doi: \begingroup \urlstyle{rm}\Url}\fi

\bibitem[Abebe et~al.(2020)Abebe, Barocas, Kleinberg, Levy, Raghavan, and Robinson]{abebe_roles_2020}
Abebe, R., Barocas, S., Kleinberg, J., Levy, K., Raghavan, M., and Robinson, D.~G.
\newblock Roles for computing in social change.
\newblock In \emph{Proceedings of the 2020 {Conference} on {Fairness}, {Accountability}, and {Transparency}}, pp.\  252--260, Barcelona Spain, January 2020. ACM.
\newblock ISBN 978-1-4503-6936-7.
\newblock \doi{10.1145/3351095.3372871}.
\newblock URL \url{https://dl.acm.org/doi/10.1145/3351095.3372871}.

\bibitem[Adams et~al.(2012)Adams, Arel, Bach, Coop, Furlan, Goertzel, Hall, Samsonovich, Scheutz, Schlesinger, et~al.]{adams2012mapping}
Adams, S., Arel, I., Bach, J., Coop, R., Furlan, R., Goertzel, B., Hall, J.~S., Samsonovich, A., Scheutz, M., Schlesinger, M., et~al.
\newblock Mapping the landscape of human-level artificial general intelligence.
\newblock \emph{AI magazine}, 33\penalty0 (1):\penalty0 25--42, 2012.

\bibitem[Agrawal et~al.(2022)Agrawal, Hegselmann, Lang, Kim, and Sontag]{agrawal_large_2022}
Agrawal, M., Hegselmann, S., Lang, H., Kim, Y., and Sontag, D.
\newblock Large language models are few-shot clinical information extractors.
\newblock In Goldberg, Y., Kozareva, Z., and Zhang, Y. (eds.), \emph{Proceedings of the 2022 {Conference} on {Empirical} {Methods} in {Natural} {Language} {Processing}}, pp.\  1998--2022, Abu Dhabi, United Arab Emirates, December 2022. Association for Computational Linguistics.
\newblock \doi{10.18653/v1/2022.emnlp-main.130}.
\newblock URL \url{https://aclanthology.org/2022.emnlp-main.130/}.

\bibitem[Agre(2014)]{agre2014toward}
Agre, P.~E.
\newblock Toward a critical technical practice: Lessons learned in trying to reform {AI}.
\newblock In \emph{Social science, technical systems, and cooperative work}, pp.\  131--157. Psychology Press, 2014.

\bibitem[Agüera~y Arcas \& Norvig(2023)Agüera~y Arcas and Norvig]{aguera_y_arcas_artificial_2023}
Agüera~y Arcas, B. and Norvig, P.
\newblock Artificial {General} {Intelligence} {Is} {Already} {Here}, October 2023.
\newblock URL \url{https://www.noemamag.com/artificial-general-intelligence-is-already-here}.

\bibitem[Ahmed et~al.(2024)Ahmed, Jaźwińska, Ahlawat, Winecoff, and Wang]{ahmed_field-building_2024}
Ahmed, S., Jaźwińska, K., Ahlawat, A., Winecoff, A., and Wang, M.
\newblock Field-building and the epistemic culture of {AI} safety.
\newblock \emph{First Monday}, April 2024.
\newblock ISSN 1396-0466.
\newblock \doi{20240428092345000}.
\newblock URL \url{https://firstmonday.org/ojs/index.php/fm/article/view/13626}.

\bibitem[Alexandrova \& Fabian(2022)Alexandrova and Fabian]{alexandrova_democratising_2022}
Alexandrova, A. and Fabian, M.
\newblock Democratising {Measurement}: or {Why} {Thick} {Concepts} {Call} for {Coproduction}.
\newblock \emph{European Journal for Philosophy of Science}, 12\penalty0 (1):\penalty0 7, January 2022.
\newblock ISSN 1879-4920.
\newblock \doi{10.1007/s13194-021-00437-7}.
\newblock URL \url{https://doi.org/10.1007/s13194-021-00437-7}.

\bibitem[Allen et~al.(2019)Allen, O’Connell, and Kiermer]{allen-etal-2019}
Allen, L., O’Connell, A., and Kiermer, V.
\newblock How can we ensure visibility and diversity in research contributions? how the contributor role taxonomy ({CRediT}) is helping the shift from authorship to contributorship.
\newblock \emph{Learned Publishing}, 32\penalty0 (1):\penalty0 71--74, 2019.
\newblock \doi{https://doi.org/10.1002/leap.1210}.
\newblock URL \url{https://onlinelibrary.wiley.com/doi/abs/10.1002/leap.1210}.

\bibitem[Altmeyer et~al.(2024)Altmeyer, Demetriou, Bartlett, and Liem]{altmeyer_position_2024}
Altmeyer, P., Demetriou, A.~M., Bartlett, A., and Liem, C. C.~S.
\newblock Position: {Stop} {Making} {Unscientific} {AGI} {Performance} {Claims}.
\newblock In \emph{Proceedings of the 41st {International} {Conference} on {Machine} {Learning}}, pp.\  1222--1242. PMLR, July 2024.
\newblock URL \url{https://proceedings.mlr.press/v235/altmeyer24a.html}.
\newblock ISSN: 2640-3498.

\bibitem[Alvarez(2023)]{sep-reasons-just-vs-expl}
Alvarez, M.
\newblock {Reasons for Action: Justification, Motivation, Explanation}.
\newblock In Zalta, E.~N. and Nodelman, U. (eds.), \emph{The {Stanford} Encyclopedia of Philosophy}. Metaphysics Research Lab, Stanford University, {W}inter 2023 edition, 2023.

\bibitem[Amodei(2024)]{machines-grace}
Amodei, D.
\newblock Machines of loving grace, 2024.
\newblock URL \url{https://www.darioamodei.com/essay/machines-of-loving-grace}.
\newblock [Online; accessed 19-May-2025].

\bibitem[Amoo et~al.(2020)Amoo, Bringardner, Chen, Coyle, Finnegan, Kim, Koman, Lagoudas, Llewellyn, Logan, et~al.]{amoo2020breaking}
Amoo, M.~E., Bringardner, J., Chen, J.-Y., Coyle, E.~J., Finnegan, J., Kim, C.~J., Koman, P.~D., Lagoudas, M.~Z., Llewellyn, D.~C., Logan, L., et~al.
\newblock Breaking down the silos: Innovations for multidisciplinary programs.
\newblock In \emph{2020 ASEE Virtual Annual Conference Content Access}, 2020.

\bibitem[Anderson(2002)]{anderson_situated_2002}
Anderson, E.
\newblock Situated {Knowledge} and the {Interplay} of {Value} {Judgments} and {Evidence} in {Scientific} {Inquiry}.
\newblock In Gärdenfors, P., Woleński, J., and Kijania-Placek, K. (eds.), \emph{In the {Scope} of {Logic}, {Methodology} and {Philosophy} of {Science}: {Volume} {Two} of the 11th {International} {Congress} of {Logic}, {Methodology} and {Philosophy} of {Science}, {Cracow}, {August} 1999}, Synthese {Library}, pp.\  497--517. Springer Netherlands, Dordrecht, 2002.
\newblock ISBN 978-94-017-0475-5.
\newblock \doi{10.1007/978-94-017-0475-5_8}.
\newblock URL \url{https://doi.org/10.1007/978-94-017-0475-5_8}.

\bibitem[Anderson(2006)]{anderson_epistemology_2006}
Anderson, E.
\newblock The {Epistemology} of {Democracy}.
\newblock \emph{Episteme}, 3\penalty0 (1-2):\penalty0 8--22, June 2006.
\newblock ISSN 1750-0117, 1742-3600.
\newblock \doi{10.3366/epi.2006.3.1-2.8}.
\newblock URL \url{https://www.cambridge.org/core/journals/episteme/article/abs/epistemology-of-democracy/F86F1D124D2E081116611043BD54CBD9}.

\bibitem[Andrews et~al.(2024)Andrews, Smart, and Birhane]{andrews_reanimation_2024}
Andrews, M., Smart, A., and Birhane, A.
\newblock The reanimation of pseudoscience in machine learning and its ethical repercussions.
\newblock \emph{Patterns}, 0\penalty0 (0), August 2024.
\newblock ISSN 2666-3899.
\newblock \doi{10.1016/j.patter.2024.101027}.
\newblock URL \url{https://www.cell.com/patterns/abstract/S2666-3899(24)00160-0}.

\bibitem[Anthropic(2025)]{anthropic-ostp}
Anthropic.
\newblock Anthropic's recommendations to {OSTP} for the {U.S. AI Action Plan}, 2025.
\newblock URL \url{https://www.anthropic.com/news/anthropic-s-recommendations-ostp-u-s-ai-action-plan}.
\newblock [Online; accessed 19-May-2025].

\bibitem[Appelman(2023)]{racismandtechnologyRacistTechnology}
Appelman, N.
\newblock {R}acist {T}echnology in {A}ction: {I}mage recognition is still not capable of differentiating gorillas from {B}lack people --- racismandtechnology.center.
\newblock \url{https://racismandtechnology.center/2023/06/09/racist-technology-in-action-image-recognition-is-still-not-capable-of-differentiating-gorillas-from-black-people/}, 2023.
\newblock [Accessed 24-01-2025].

\bibitem[Attard-Frost(2023)]{attard-frost_queering_2023}
Attard-Frost, B.
\newblock Queering intelligence: {A} theory of intelligence as performance and a critique of individual and artificial intelligence.
\newblock In \emph{Queer {Reflections} on {AI}}. Routledge, 2023.
\newblock ISBN 978-1-00-335795-7.
\newblock URL \url{https://www.taylorfrancis.com/chapters/oa-edit/10.4324/9781003357957-3/queering-intelligence-blair-attard-frost}.

\bibitem[Ballantyne(2019)]{ballantyne_epistemic_2019}
Ballantyne, N.
\newblock Epistemic {Trespassing}.
\newblock \emph{Mind}, 128\penalty0 (510):\penalty0 367--395, April 2019.
\newblock ISSN 0026-4423.
\newblock \doi{10.1093/mind/fzx042}.
\newblock URL \url{https://doi.org/10.1093/mind/fzx042}.

\bibitem[Balloccu et~al.(2024)Balloccu, Schmidtov{\'a}, Lango, and Dusek]{balloccu-etal-2024-leak}
Balloccu, S., Schmidtov{\'a}, P., Lango, M., and Dusek, O.
\newblock Leak, cheat, repeat: Data contamination and evaluation malpractices in closed-source {LLM}s.
\newblock In Graham, Y. and Purver, M. (eds.), \emph{Proceedings of the 18th Conference of the European Chapter of the Association for Computational Linguistics (Volume 1: Long Papers)}, pp.\  67--93, St. Julian{'}s, Malta, March 2024. Association for Computational Linguistics.
\newblock URL \url{https://aclanthology.org/2024.eacl-long.5/}.

\bibitem[Bauer \& Gill(2024)Bauer and Gill]{bauer_mirror_2024}
Bauer, K. and Gill, A.
\newblock Mirror, {Mirror} on the {Wall}: {Algorithmic} {Assessments}, {Transparency}, and {Self}-{Fulfilling} {Prophecies}.
\newblock \emph{Information Systems Research}, 35\penalty0 (1):\penalty0 226--248, March 2024.
\newblock ISSN 1047-7047.
\newblock \doi{10.1287/isre.2023.1217}.
\newblock URL \url{https://pubsonline.informs.org/doi/full/10.1287/isre.2023.1217}.

\bibitem[{BBC News}(2015)]{bbcGoogleApologises}
{BBC News}.
\newblock {G}oogle apologises for {P}hotos app's racist blunder.
\newblock \url{https://www.bbc.com/news/technology-33347866}, 2015.
\newblock [Accessed 24-01-2025].

\bibitem[Bender et~al.(2021)Bender, Gebru, McMillan-Major, and Shmitchell]{bender_dangers_2021}
Bender, E.~M., Gebru, T., McMillan-Major, A., and Shmitchell, S.
\newblock On the {Dangers} of {Stochastic} {Parrots}: {Can} {Language} {Models} {Be} {Too} {Big}?
\newblock In \emph{Proceedings of the 2021 {ACM} {Conference} on {Fairness}, {Accountability}, and {Transparency}}, {FAccT} '21, pp.\  610--623, New York, NY, USA, March 2021. Association for Computing Machinery.
\newblock ISBN 978-1-4503-8309-7.
\newblock \doi{10.1145/3442188.3445922}.
\newblock URL \url{https://doi.org/10.1145/3442188.3445922}.

\bibitem[Bergman et~al.(2024)Bergman, Marchal, Mellor, Mohamed, Gabriel, and Isaac]{bergman_stela_2024}
Bergman, S., Marchal, N., Mellor, J., Mohamed, S., Gabriel, I., and Isaac, W.
\newblock {STELA}: a community-centred approach to norm elicitation for {AI} alignment.
\newblock \emph{Scientific Reports}, 14\penalty0 (1):\penalty0 6616, March 2024.
\newblock ISSN 2045-2322.
\newblock \doi{10.1038/s41598-024-56648-4}.
\newblock URL \url{https://www.nature.com/articles/s41598-024-56648-4}.
\newblock Publisher: Nature Publishing Group.

\bibitem[Bertelsen et~al.(2024)Bertelsen, Bossen, Knudsen, and Pedersen]{bertelsen_data_2024}
Bertelsen, P.~S., Bossen, C., Knudsen, C., and Pedersen, A.~M.
\newblock Data work and practices in healthcare: {A} scoping review.
\newblock \emph{International Journal of Medical Informatics}, 184:\penalty0 105348, April 2024.
\newblock ISSN 1386-5056.
\newblock \doi{10.1016/j.ijmedinf.2024.105348}.
\newblock URL \url{https://www.sciencedirect.com/science/article/pii/S138650562400011X}.

\bibitem[Binns(2018)]{binns_algorithmic_2018}
Binns, R.
\newblock Algorithmic {Accountability} and {Public} {Reason}.
\newblock \emph{Philosophy \& Technology}, 31\penalty0 (4):\penalty0 543--556, December 2018.
\newblock ISSN 2210-5433, 2210-5441.
\newblock \doi{10.1007/s13347-017-0263-5}.
\newblock URL \url{http://link.springer.com/10.1007/s13347-017-0263-5}.

\bibitem[Birhane \& Guest(2021)Birhane and Guest]{birhane_towards_2021}
Birhane, A. and Guest, O.
\newblock Towards {Decolonising} {Computational} {Sciences}.
\newblock \emph{Kvinder, Køn \& Forskning}, 2021.
\newblock \doi{10.7146/kkf.v29i2.124899}.
\newblock URL \url{https://pure.mpg.de/rest/items/item_3287104_1/component/file_3287105/content}.

\bibitem[Birhane et~al.(2022{\natexlab{a}})Birhane, Isaac, Prabhakaran, Diaz, Elish, Gabriel, and Mohamed]{birhane_power_2022}
Birhane, A., Isaac, W., Prabhakaran, V., Diaz, M., Elish, M.~C., Gabriel, I., and Mohamed, S.
\newblock Power to the {People}? {Opportunities} and {Challenges} for {Participatory} {AI}.
\newblock In \emph{Proceedings of the 2nd {ACM} {Conference} on {Equity} and {Access} in {Algorithms}, {Mechanisms}, and {Optimization}}, {EAAMO} '22, pp.\  1--8, New York, NY, USA, October 2022{\natexlab{a}}. Association for Computing Machinery.
\newblock ISBN 978-1-4503-9477-2.
\newblock \doi{10.1145/3551624.3555290}.
\newblock URL \url{https://dl.acm.org/doi/10.1145/3551624.3555290}.

\bibitem[Birhane et~al.(2022{\natexlab{b}})Birhane, Kalluri, Card, Agnew, Dotan, and Bao]{birhane_values_2022}
Birhane, A., Kalluri, P., Card, D., Agnew, W., Dotan, R., and Bao, M.
\newblock The {Values} {Encoded} in {Machine} {Learning} {Research}.
\newblock In \emph{2022 {ACM} {Conference} on {Fairness}, {Accountability}, and {Transparency}}, pp.\  173--184, Seoul Republic of Korea, June 2022{\natexlab{b}}. ACM.
\newblock ISBN 978-1-4503-9352-2.
\newblock \doi{10.1145/3531146.3533083}.
\newblock URL \url{https://dl.acm.org/doi/10.1145/3531146.3533083}.

\bibitem[Blili-Hamelin \& Hancox-Li(2023)Blili-Hamelin and Hancox-Li]{blili-hamelin_making_2023}
Blili-Hamelin, B. and Hancox-Li, L.
\newblock Making {Intelligence}: {Ethical} {Values} in {IQ} and {ML} {Benchmarks}.
\newblock In \emph{Proceedings of the 2023 {ACM} {Conference} on {Fairness}, {Accountability}, and {Transparency}}, {FAccT} '23, pp.\  271--284, New York, NY, USA, June 2023. Association for Computing Machinery.
\newblock ISBN 9798400701924.
\newblock \doi{10.1145/3593013.3593996}.
\newblock URL \url{https://dl.acm.org/doi/10.1145/3593013.3593996}.

\bibitem[Blili-Hamelin et~al.(2024)Blili-Hamelin, Hancox-Li, and Smart]{blili-hamelin_unsocial_2024}
Blili-Hamelin, B., Hancox-Li, L., and Smart, A.
\newblock Unsocial {Intelligence}: {An} {Investigation} of the {Assumptions} of {AGI} {Discourse}.
\newblock \emph{Proceedings of the AAAI/ACM Conference on AI, Ethics, and Society}, 7:\penalty0 141--155, October 2024.
\newblock URL \url{https://ojs.aaai.org/index.php/AIES/article/view/31625}.

\bibitem[Blodgett et~al.(2020)Blodgett, Barocas, Daumé~Iii, and Wallach]{blodgett_language_2020}
Blodgett, S.~L., Barocas, S., Daumé~Iii, H., and Wallach, H.
\newblock Language ({Technology}) is {Power}: {A} {Critical} {Survey} of “{Bias}” in {NLP}.
\newblock In \emph{Proceedings of the 58th {Annual} {Meeting} of the {Association} for {Computational} {Linguistics}}, pp.\  5454--5476, Online, 2020. Association for Computational Linguistics.
\newblock \doi{10.18653/v1/2020.acl-main.485}.
\newblock URL \url{https://www.aclweb.org/anthology/2020.acl-main.485}.

\bibitem[Bommasani(2023)]{bommasani_evaluation_2023}
Bommasani, R.
\newblock Evaluation for {Change}.
\newblock In Rogers, A., Boyd-Graber, J., and Okazaki, N. (eds.), \emph{Findings of the {Association} for {Computational} {Linguistics}: {ACL} 2023}, pp.\  8227--8239, Toronto, Canada, July 2023. Association for Computational Linguistics.
\newblock \doi{10.18653/v1/2023.findings-acl.522}.
\newblock URL \url{https://aclanthology.org/2023.findings-acl.522/}.

\bibitem[Bommasani et~al.(2022)Bommasani, Creel, Kumar, Jurafsky, and Liang]{bommasani_picking_2022}
Bommasani, R., Creel, K.~A., Kumar, A., Jurafsky, D., and Liang, P.~S.
\newblock Picking on the {Same} {Person}: {Does} {Algorithmic} {Monoculture} lead to {Outcome} {Homogenization}?
\newblock \emph{Advances in Neural Information Processing Systems}, 35:\penalty0 3663--3678, December 2022.
\newblock URL \url{https://proceedings.neurips.cc/paper_files/paper/2022/hash/17a234c91f746d9625a75cf8a8731ee2-Abstract-Conference.html}.

\bibitem[Bommasani et~al.(2025)Bommasani, Singer, Appel, Cen, Cooper, Cryst, Gailmard, Gonzalez, Ho, Klaus, Lee, Liang, Reuel, Song, Spence, Wan, Wang, Zhang, Zittrain, Tour~Chayes, Cuéllar, and Fei-Fei]{rishi_bommasani_draft_2025}
Bommasani, R., Singer, S., Appel, R.~E., Cen, S., Cooper, A.~F., Cryst, E., Gailmard, L.~A., Gonzalez, J.~E., Ho, D.~E., Klaus, I., Lee, M.~M., Liang, P., Reuel, A., Song, D., Spence, D., Wan, A., Wang, A., Zhang, D., Zittrain, J., Tour~Chayes, J., Cuéllar, M.-F., and Fei-Fei, L.
\newblock {DRAFT} {REPORT} of the {Joint} {California} {Policy} {Working} {Group} on {AI} {Frontier} {Models}.
\newblock Technical report, Joint California Policy Working Group on AI Frontier Models, March 2025.
\newblock URL \url{https://www.cafrontieraigov.org/wp-content/uploads/2025/03/Draft_Report_of_the_Joint_California_Policy_Working_Group_on_AI_Frontier_Models.pdf}.

\bibitem[Bostrom(2014)]{bostrom_superintelligence_2014}
Bostrom, N.
\newblock \emph{Superintelligence: {Paths}, dangers, strategies}.
\newblock Oxford University Press, New York, NY, US, 2014.
\newblock ISBN 978-0-19-967811-2.

\bibitem[Bouthillier et~al.(2019)Bouthillier, Laurent, and Vincent]{pmlr-v97-bouthillier19a}
Bouthillier, X., Laurent, C., and Vincent, P.
\newblock Unreproducible research is reproducible.
\newblock In Chaudhuri, K. and Salakhutdinov, R. (eds.), \emph{Proceedings of the 36th International Conference on Machine Learning}, volume~97 of \emph{Proceedings of Machine Learning Research}, pp.\  725--734. PMLR, 09--15 Jun 2019.
\newblock URL \url{https://proceedings.mlr.press/v97/bouthillier19a.html}.

\bibitem[Broussard et~al.(2019)Broussard, Diakopoulos, Guzman, Abebe, Dupagne, and Chuan]{broussard_artificial_2019}
Broussard, M., Diakopoulos, N., Guzman, A.~L., Abebe, R., Dupagne, M., and Chuan, C.-H.
\newblock Artificial {Intelligence} and {Journalism}.
\newblock \emph{Journalism \& Mass Communication Quarterly}, 96\penalty0 (3):\penalty0 673--695, September 2019.
\newblock ISSN 1077-6990, 2161-430X.
\newblock \doi{10.1177/1077699019859901}.
\newblock URL \url{http://journals.sagepub.com/doi/10.1177/1077699019859901}.

\bibitem[Browne(2025)]{CNBC-AGI}
Browne, R.
\newblock {AI} that can match humans at any task will be here in five to 10 years, {Google DeepMind CEO} says.
\newblock \url{https://www.cnbc.com/2025/03/17/human-level-ai-will-be-here-in-5-to-10-years-deepmind-ceo-says.html}, 2025.

\bibitem[Bubeck et~al.(2023)Bubeck, Chandrasekaran, Eldan, Gehrke, Horvitz, Kamar, Lee, Lee, Li, Lundberg, Nori, Palangi, Ribeiro, and Zhang]{bubeck2023sparksartificialgeneralintelligence}
Bubeck, S., Chandrasekaran, V., Eldan, R., Gehrke, J., Horvitz, E., Kamar, E., Lee, P., Lee, Y.~T., Li, Y., Lundberg, S., Nori, H., Palangi, H., Ribeiro, M.~T., and Zhang, Y.
\newblock Sparks of artificial general intelligence: Early experiments with {GPT}-4, 2023.
\newblock URL \url{https://arxiv.org/abs/2303.12712}.

\bibitem[Bucknall \& Dori-Hacohen(2022)Bucknall and Dori-Hacohen]{bucknall_current_2022}
Bucknall, B.~S. and Dori-Hacohen, S.
\newblock Current and {Near}-{Term} {AI} as a {Potential} {Existential} {Risk} {Factor}.
\newblock In \emph{Proceedings of the 2022 {AAAI}/{ACM} {Conference} on {AI}, {Ethics}, and {Society}}, {AIES} '22, pp.\  119--129, New York, NY, USA, July 2022. Association for Computing Machinery.
\newblock ISBN 978-1-4503-9247-1.
\newblock \doi{10.1145/3514094.3534146}.
\newblock URL \url{https://dl.acm.org/doi/10.1145/3514094.3534146}.

\bibitem[Buolamwini \& Gebru(2018)Buolamwini and Gebru]{buolamwini_gender_2018}
Buolamwini, J. and Gebru, T.
\newblock Gender {Shades}: {Intersectional} {Accuracy} {Disparities} in {Commercial} {Gender} {Classification}.
\newblock In \emph{Proceedings of the 1st {Conference} on {Fairness}, {Accountability} and {Transparency}}, pp.\  77--91. PMLR, January 2018.
\newblock URL \url{https://proceedings.mlr.press/v81/buolamwini18a.html}.

\bibitem[Burt(2004)]{burt_structural_2004}
Burt, R.
\newblock Structural {Holes} and {Good} {Ideas}.
\newblock \emph{American Journal of Sociology}, 110\penalty0 (2):\penalty0 349--399, September 2004.
\newblock ISSN 0002-9602.
\newblock \doi{10.1086/421787}.
\newblock URL \url{https://www.journals.uchicago.edu/doi/full/10.1086/421787}.
\newblock Publisher: The University of Chicago Press.

\bibitem[Calin-Jageman \& Cumming(2019)Calin-Jageman and Cumming]{calin2019new}
Calin-Jageman, R.~J. and Cumming, G.
\newblock The new statistics for better science: Ask how much, how uncertain, and what else is known.
\newblock \emph{The American Statistician}, 73\penalty0 (sup1):\penalty0 271--280, 2019.

\bibitem[Cameron(2023)]{cameron_us_2023}
Cameron, D.
\newblock {US} {Justice} {Department} {Urged} to {Investigate} {Gunshot} {Detector} {Purchases}.
\newblock \emph{Wired}, September 2023.
\newblock ISSN 1059-1028.
\newblock URL \url{https://www.wired.com/story/shotspotter-doj-letter-epic/}.

\bibitem[Cao(2022)]{cao_finance2022}
Cao, L.
\newblock Ai in finance: Challenges, techniques, and opportunities.
\newblock \emph{ACM Computing Surveys}, 55\penalty0 (3), February 2022.
\newblock ISSN 0360-0300.
\newblock \doi{10.1145/3502289}.
\newblock URL \url{https://doi.org/10.1145/3502289}.

\bibitem[Cave(2020)]{cave_problem_2020}
Cave, S.
\newblock The {Problem} with {Intelligence}: {Its} {Value}-{Laden} {History} and the {Future} of {AI}.
\newblock In \emph{Proceedings of the {AAAI}/{ACM} {Conference} on {AI}, {Ethics}, and {Society}}, {AIES} '20, pp.\  29--35, New York, NY, USA, February 2020. Association for Computing Machinery.
\newblock ISBN 978-1-4503-7110-0.
\newblock \doi{10.1145/3375627.3375813}.
\newblock URL \url{https://dl.acm.org/doi/10.1145/3375627.3375813}.

\bibitem[Chalmers(2010)]{chalmers_singularity_2010}
Chalmers, D.~J.
\newblock The singularity: {A} philosophical analysis.
\newblock \emph{Journal of Consciousness Studies}, 17\penalty0 (9-10):\penalty0 9 -- 10, 2010.

\bibitem[Chollet(2019)]{chollet_measure_2019}
Chollet, F.
\newblock On the {Measure} of {Intelligence}, November 2019.
\newblock URL \url{http://arxiv.org/abs/1911.01547}.

\bibitem[Chollet(2024{\natexlab{a}})]{chollet_openai_2024}
Chollet, F.
\newblock {OpenAI} o3 {Breakthrough} {High} {Score} on {ARC}-{AGI}-{Pub}, December 2024{\natexlab{a}}.
\newblock URL \url{https://arcprize.org/blog/oai-o3-pub-breakthrough}.

\bibitem[Chollet(2024{\natexlab{b}})]{chollet_so_2024}
Chollet, F.
\newblock "{So}, is this {AGI}?...", December 2024{\natexlab{b}}.
\newblock URL \url{https://x.com/fchollet/status/1870170778458828851}.

\bibitem[Chollet et~al.(2024)Chollet, Knoop, Kamradt, and Landers]{chollet_arc_technical_report_2024}
Chollet, F., Knoop, M., Kamradt, G., and Landers, B.
\newblock {ARC} {Prize} 2024: {Technical} {Report}.
\newblock Technical report, ARC-AGI, December 2024.

\bibitem[Church \& Kordoni(2022)Church and Kordoni]{church_emerging_2022}
Church, K.~W. and Kordoni, V.
\newblock Emerging {Trends}: {SOTA}-{Chasing}.
\newblock \emph{Natural Language Engineering}, 28\penalty0 (2):\penalty0 249--269, March 2022.
\newblock ISSN 1351-3249, 1469-8110.
\newblock \doi{10.1017/S1351324922000043}.
\newblock URL \url{https://www.cambridge.org/core/product/identifier/S1351324922000043/type/journal_article}.

\bibitem[Connealy et~al.(2024)Connealy, Piza, Arietti, Mohler, and Carter]{connealy_staggered_2024}
Connealy, N.~T., Piza, E.~L., Arietti, R.~A., Mohler, G.~O., and Carter, J.~G.
\newblock Staggered deployment of gunshot detection technology in {Chicago}, {IL}: a matched quasi-experiment of gun violence outcomes.
\newblock \emph{Journal of Experimental Criminology}, March 2024.
\newblock ISSN 1572-8315.
\newblock \doi{10.1007/s11292-024-09617-w}.
\newblock URL \url{https://doi.org/10.1007/s11292-024-09617-w}.

\bibitem[Cooper et~al.(2022)Cooper, Horne, Hayes, Heldreth, Lahav, Holbrook, and Wilcox]{cooper_systematic_2022}
Cooper, N., Horne, T., Hayes, G.~R., Heldreth, C., Lahav, M., Holbrook, J., and Wilcox, L.
\newblock A {Systematic} {Review} and {Thematic} {Analysis} of {Community}-{Collaborative} {Approaches} to {Computing} {Research}.
\newblock In \emph{Proceedings of the 2022 {CHI} {Conference} on {Human} {Factors} in {Computing} {Systems}}, {CHI} '22, pp.\  1--18, New York, NY, USA, April 2022. Association for Computing Machinery.
\newblock ISBN 978-1-4503-9157-3.
\newblock \doi{10.1145/3491102.3517716}.
\newblock URL \url{https://dl.acm.org/doi/10.1145/3491102.3517716}.

\bibitem[Costanza-Chock(2020)]{costanza-chock_design_2020}
Costanza-Chock, S.
\newblock \emph{Design {Justice}: {Community}-{Led} {Practices} to {Build} the {Worlds} {We} {Need}}.
\newblock The MIT Press, 2020.
\newblock ISBN 978-0-262-04345-8.
\newblock URL \url{https://library.oapen.org/handle/20.500.12657/43542}.

\bibitem[D'Amour et~al.(2022)D'Amour, Heller, Moldovan, Adlam, Alipanahi, Beutel, Chen, Deaton, Eisenstein, Hoffman, Hormozdiari, Houlsby, Hou, Jerfel, Karthikesalingam, Lucic, Ma, McLean, Mincu, Mitani, Montanari, Nado, Natarajan, Nielson, Osborne, Raman, Ramasamy, Sayres, Schrouff, Seneviratne, Sequeira, Suresh, Veitch, Vladymyrov, Wang, Webster, Yadlowsky, Yun, Zhai, and Sculley]{JMLR:v23:20-1335}
D'Amour, A., Heller, K., Moldovan, D., Adlam, B., Alipanahi, B., Beutel, A., Chen, C., Deaton, J., Eisenstein, J., Hoffman, M.~D., Hormozdiari, F., Houlsby, N., Hou, S., Jerfel, G., Karthikesalingam, A., Lucic, M., Ma, Y., McLean, C., Mincu, D., Mitani, A., Montanari, A., Nado, Z., Natarajan, V., Nielson, C., Osborne, T.~F., Raman, R., Ramasamy, K., Sayres, R., Schrouff, J., Seneviratne, M., Sequeira, S., Suresh, H., Veitch, V., Vladymyrov, M., Wang, X., Webster, K., Yadlowsky, S., Yun, T., Zhai, X., and Sculley, D.
\newblock Underspecification presents challenges for credibility in modern machine learning.
\newblock \emph{Journal of Machine Learning Research}, 23\penalty0 (226):\penalty0 1--61, 2022.
\newblock URL \url{http://jmlr.org/papers/v23/20-1335.html}.

\bibitem[DeepMind(2025)]{DeepMindAbout2025}
DeepMind, G.
\newblock About, 2025.
\newblock URL \url{https://deepmind.google/about/}.
\newblock [Online; accessed 19-May-2025].

\bibitem[Dehghani et~al.(2021)Dehghani, Tay, Gritsenko, Zhao, Houlsby, Diaz, Metzler, and Vinyals]{dehghani_benchmark_2021}
Dehghani, M., Tay, Y., Gritsenko, A.~A., Zhao, Z., Houlsby, N., Diaz, F., Metzler, D., and Vinyals, O.
\newblock The {Benchmark} {Lottery}, July 2021.
\newblock URL \url{http://arxiv.org/abs/2107.07002}.

\bibitem[Delgado et~al.(2023)Delgado, Yang, Madaio, and Yang]{delgado_participatory_2023}
Delgado, F., Yang, S., Madaio, M., and Yang, Q.
\newblock The {Participatory} {Turn} in {AI} {Design}: {Theoretical} {Foundations} and the {Current} {State} of {Practice}.
\newblock In \emph{Equity and {Access} in {Algorithms}, {Mechanisms}, and {Optimization}}, pp.\  1--23, Boston MA USA, October 2023. ACM.
\newblock ISBN 9798400703812.
\newblock \doi{10.1145/3617694.3623261}.
\newblock URL \url{https://dl.acm.org/doi/10.1145/3617694.3623261}.

\bibitem[Denton et~al.(2020)Denton, Hanna, Amironesei, Smart, Nicole, and Scheuerman]{denton_bringing_2020}
Denton, E., Hanna, A., Amironesei, R., Smart, A., Nicole, H., and Scheuerman, M.~K.
\newblock Bringing the {People} {Back} {In}: {Contesting} {Benchmark} {Machine} {Learning} {Datasets}, July 2020.
\newblock URL \url{http://arxiv.org/abs/2007.07399}.
\newblock arXiv:2007.07399 [cs].

\bibitem[Denton et~al.(2021)Denton, Hanna, Amironesei, Smart, and Nicole]{denton2021genealogy}
Denton, E., Hanna, A., Amironesei, R., Smart, A., and Nicole, H.
\newblock On the genealogy of machine learning datasets: A critical history of {ImageNet}.
\newblock \emph{Big Data \& Society}, 8\penalty0 (2), 2021.

\bibitem[Devlin et~al.(2019)Devlin, Chang, Lee, and Toutanova]{devlin2019bert}
Devlin, J., Chang, M.-W., Lee, K., and Toutanova, K.
\newblock {BERT}: Pre-training of deep bidirectional transformers for language understanding.
\newblock In Burstein, J., Doran, C., and Solorio, T. (eds.), \emph{Proceedings of the 2019 Conference of the North {A}merican Chapter of the Association for Computational Linguistics: Human Language Technologies, Volume 1 (Long and Short Papers)}, pp.\  4171--4186, Minneapolis, Minnesota, June 2019. Association for Computational Linguistics.
\newblock \doi{10.18653/v1/N19-1423}.
\newblock URL \url{https://aclanthology.org/N19-1423/}.

\bibitem[DiPaolo(2022)]{dipaolo_whats_2022}
DiPaolo, J.
\newblock What’s wrong with epistemic trespassing?
\newblock \emph{Philosophical Studies}, 179\penalty0 (1):\penalty0 223--243, January 2022.
\newblock ISSN 1573-0883.
\newblock \doi{10.1007/s11098-021-01657-6}.
\newblock URL \url{https://doi.org/10.1007/s11098-021-01657-6}.

\bibitem[Dori-Hacohen et~al.(2021)Dori-Hacohen, Montenegro, Murai, Hale, Sung, Blain, and Edwards-Johnson]{dorihacohen2021fairness}
Dori-Hacohen, S., Montenegro, R.~E., Murai, F., Hale, S.~A., Sung, K., Blain, M., and Edwards-Johnson, J.
\newblock {Fairness via AI: Bias Reduction in Medical Information}.
\newblock In \emph{The 4th FAccTRec Workshop on Responsible Recommendation at RecSys}, 2021.

\bibitem[Dotan \& Milli(2020)Dotan and Milli]{dotan_value-laden_2020}
Dotan, R. and Milli, S.
\newblock Value-laden disciplinary shifts in machine learning {\textbar} {Proceedings} of the 2020 {Conference} on {Fairness}, {Accountability}, and {Transparency}.
\newblock \emph{FAT* '20: Proceedings of the 2020 Conference on Fairness, Accountability, and Transparency}, January 2020.
\newblock \doi{10.1145/3351095.3373157}.
\newblock URL \url{https://dl.acm.org/doi/abs/10.1145/3351095.3373157}.

\bibitem[Doucette et~al.(2021)Doucette, Green, Necci~Dineen, Shapiro, and Raissian]{doucette_impact_2021}
Doucette, M.~L., Green, C., Necci~Dineen, J., Shapiro, D., and Raissian, K.~M.
\newblock Impact of {ShotSpotter} {Technology} on {Firearm} {Homicides} and {Arrests} {Among} {Large} {Metropolitan} {Counties}: a {Longitudinal} {Analysis}, 1999–2016.
\newblock \emph{Journal of Urban Health : Bulletin of the New York Academy of Medicine}, 98\penalty0 (5):\penalty0 609--621, October 2021.
\newblock ISSN 1099-3460.
\newblock \doi{10.1007/s11524-021-00515-4}.
\newblock URL \url{https://www.ncbi.nlm.nih.gov/pmc/articles/PMC8566613/}.

\bibitem[Dreyfus \& Dreyfus(1986)Dreyfus and Dreyfus]{dreyfus_mind_1986}
Dreyfus, H. and Dreyfus, S.~E.
\newblock \emph{Mind {Over} {Machine}}.
\newblock Simon and Schuster, 1986.
\newblock ISBN 978-0-7432-0551-1.

\bibitem[Dulka(2022)]{dulka_use_2022}
Dulka, A.
\newblock The {Use} of {Artificial} {Intelligence} in {International} {Human} {Rights} {Law}.
\newblock \emph{Stanford Technology Law Review}, 26\penalty0 (2):\penalty0 316--366, 2022.
\newblock URL \url{https://heinonline.org/HOL/P?h=hein.journals/stantlr26&i=316}.

\bibitem[El-Mhamdi et~al.(2021)El-Mhamdi, Farhadkhani, Guerraoui, Guirguis, Hoang, and Rouault]{el2021collaborative}
El-Mhamdi, E.~M., Farhadkhani, S., Guerraoui, R., Guirguis, A., Hoang, L.-N., and Rouault, S.
\newblock Collaborative learning in the jungle (decentralized, byzantine, heterogeneous, asynchronous and nonconvex learning).
\newblock \emph{Advances in neural information processing systems}, 34:\penalty0 25044--25057, 2021.

\bibitem[El-Mhamdi et~al.(2023)El-Mhamdi, Farhadkhani, Guerraoui, Gupta, Hoang, Pinot, Rouault, and Stephan]{el2023impossible}
El-Mhamdi, E.-M., Farhadkhani, S., Guerraoui, R., Gupta, N., Hoang, L.-N., Pinot, R., Rouault, S., and Stephan, J.
\newblock On the {Impossible} {Safety} of {Large} {AI} {Models}, May 2023.
\newblock URL \url{http://arxiv.org/abs/2209.15259}.
\newblock arXiv:2209.15259 [cs].

\bibitem[{Fast Company}(2010)]{Wozniak_coffee_2010}
{Fast Company}.
\newblock Wozniak: Could a computer make a cup of coffee?, 2010.
\newblock URL \url{https://www.youtube.com/watch?v=MowergwQR5Y}.
\newblock [Online; accessed 17-January-2023].

\bibitem[Fei et~al.(2022)Fei, Lu, Gao, Yang, Huo, Wen, Lu, Song, Gao, Xiang, Sun, and Wen]{fei_towards_2022}
Fei, N., Lu, Z., Gao, Y., Yang, G., Huo, Y., Wen, J., Lu, H., Song, R., Gao, X., Xiang, T., Sun, H., and Wen, J.-R.
\newblock Towards artificial general intelligence via a multimodal foundation model.
\newblock \emph{Nature Communications}, 13\penalty0 (1):\penalty0 3094, June 2022.
\newblock ISSN 2041-1723.
\newblock \doi{10.1038/s41467-022-30761-2}.
\newblock URL \url{https://www.nature.com/articles/s41467-022-30761-2}.
\newblock Publisher: Nature Publishing Group.

\bibitem[Fishman \& Hancox-Li(2022)Fishman and Hancox-Li]{fishman_should_2022}
Fishman, N. and Hancox-Li, L.
\newblock Should attention be all we need? {The} epistemic and ethical implications of unification in machine learning.
\newblock In \emph{Proceedings of the 2022 {ACM} {Conference} on {Fairness}, {Accountability}, and {Transparency}}, {FAccT} '22, pp.\  1516--1527, New York, NY, USA, June 2022. Association for Computing Machinery.
\newblock ISBN 978-1-4503-9352-2.
\newblock \doi{10.1145/3531146.3533206}.
\newblock URL \url{https://dl.acm.org/doi/10.1145/3531146.3533206}.

\bibitem[Fourcade et~al.(2015)Fourcade, Ollion, and Algan]{fourcade_superiority_2015}
Fourcade, M., Ollion, E., and Algan, Y.
\newblock The {Superiority} of {Economists}.
\newblock \emph{Journal of Economic Perspectives}, 29\penalty0 (1):\penalty0 89--114, February 2015.
\newblock ISSN 0895-3309.
\newblock \doi{10.1257/jep.29.1.89}.
\newblock URL \url{https://www.aeaweb.org/articles?id=10.1257/jep.29.1.89}.

\bibitem[{Francesca Rossi} et~al.(2025){Francesca Rossi}, {Christian Bessiere}, {Joydeep Biswas}, {Rodney Brooks}, {Vincent Conitzer}, {Thomas G. Dietterich}, {Virginia Dignum}, {Oren Etzioni}, {Kenneth D. Forbus}, {Eugene Freuder}, {Yolanda Gil}, {Holger Hoos}, {Eric Horvitz}, {Subbarao Kambhampati}, {Henry Kautz}, {Jihie Kim}, {Hiroaki Kitano}, {Alan Mackworth}, {Karen Myers}, {Luc De Raedt}, {Stuart Russell}, {Bart Selman}, {Peter Stone}, {Millind Tambe}, and {Michael Wooldridge}]{francesca_rossi_aaai_2025}
{Francesca Rossi}, {Christian Bessiere}, {Joydeep Biswas}, {Rodney Brooks}, {Vincent Conitzer}, {Thomas G. Dietterich}, {Virginia Dignum}, {Oren Etzioni}, {Kenneth D. Forbus}, {Eugene Freuder}, {Yolanda Gil}, {Holger Hoos}, {Eric Horvitz}, {Subbarao Kambhampati}, {Henry Kautz}, {Jihie Kim}, {Hiroaki Kitano}, {Alan Mackworth}, {Karen Myers}, {Luc De Raedt}, {Stuart Russell}, {Bart Selman}, {Peter Stone}, {Millind Tambe}, and {Michael Wooldridge}.
\newblock {AAAI} 2025 presidential panel on the future of {AI} research.
\newblock Technical report, Association for the Advancement of Artificial Intelligence, March 2025.
\newblock URL \url{https://aaai.org/wp-content/uploads/2025/03/AAAI-2025-PresPanel-Report-Digital-3.7.25.pdf}.

\bibitem[Frank et~al.(2017)Frank, Roehrig, and Pring]{frank2017machines}
Frank, M., Roehrig, P., and Pring, B.
\newblock \emph{What to do when machines do everything: How to get ahead in a world of {AI}, algorithms, bots, and big data}.
\newblock John Wiley \& Sons, 2017.

\bibitem[Friedler et~al.(2021)Friedler, Scheidegger, and Venkatasubramanian]{friedler_impossibility_2021}
Friedler, S.~A., Scheidegger, C., and Venkatasubramanian, S.
\newblock The ({Im})possibility of fairness: different value systems require different mechanisms for fair decision making.
\newblock \emph{Communications of the ACM}, 64\penalty0 (4):\penalty0 136--143, April 2021.
\newblock ISSN 0001-0782, 1557-7317.
\newblock \doi{10.1145/3433949}.
\newblock URL \url{https://dl.acm.org/doi/10.1145/3433949}.

\bibitem[Gabriel(2020)]{gabriel_artificial_2020}
Gabriel, I.
\newblock Artificial {Intelligence}, {Values}, and {Alignment}.
\newblock \emph{Minds and Machines}, 30\penalty0 (3):\penalty0 411--437, September 2020.
\newblock ISSN 0924-6495, 1572-8641.
\newblock \doi{10.1007/s11023-020-09539-2}.
\newblock URL \url{https://link.springer.com/10.1007/s11023-020-09539-2}.

\bibitem[Gebru \& Torres(2024)Gebru and Torres]{gebru_tescreal_2024}
Gebru, T. and Torres, E.~P.
\newblock The {TESCREAL} bundle: {Eugenics} and the promise of utopia through artificial general intelligence.
\newblock \emph{First Monday}, April 2024.
\newblock ISSN 1396-0466.
\newblock \doi{20240428092319000}.
\newblock URL \url{https://firstmonday.org/ojs/index.php/fm/article/view/13636}.

\bibitem[Goertzel(2014)]{goertzel_artificial_2014}
Goertzel, B.
\newblock Artificial general intelligence: concept, state of the art, and future prospects.
\newblock \emph{Journal of Artificial General Intelligence}, 5\penalty0 (1):\penalty0 1, 2014.
\newblock URL \url{https://sciendo.com/abstract/journals/jagi/5/1/article-p1.xml}.

\bibitem[Goertzel et~al.(2012)Goertzel, Ikl{\'e}, and Wigmore]{goertzel2012architecture}
Goertzel, B., Ikl{\'e}, M., and Wigmore, J.
\newblock The architecture of human-like general intelligence.
\newblock In \emph{Theoretical foundations of artificial general intelligence}, pp.\  123--144. Springer, 2012.

\bibitem[Gopnik(2019)]{gopnik_ais_2019}
Gopnik, A.
\newblock {AIs} {Versus} {Four}-{Year}-{Olds}.
\newblock In Brockman, J. (ed.), \emph{Possible minds: twenty-five ways of looking at {AI}}. Penguin Press, New York, 2019.
\newblock ISBN 978-0-525-55799-9 978-0-525-55801-9.

\bibitem[Gould(1981)]{gould_mismeasure_1981}
Gould, S.~J.
\newblock \emph{The mismeasure of man}.
\newblock Norton, New York, 1st ed edition, 1981.
\newblock ISBN 978-0-393-01489-1.

\bibitem[Grant \& Hill(2023)Grant and Hill]{nytimesGooglesPhoto}
Grant, N. and Hill, K.
\newblock {G}oogle’s {P}hoto {A}pp {S}till {C}an’t {F}ind {G}orillas. {A}nd {N}either {C}an {A}pple’s. ({P}ublished 2023) --- nytimes.com.
\newblock \url{https://www.nytimes.com/2023/05/22/technology/ai-photo-labels-google-apple.html}, 2023.
\newblock [Accessed 24-01-2025].

\bibitem[Graziul et~al.(2023)Graziul, Belikov, Chattopadyay, Chen, Fang, Girdhar, Jia, Krafft, Kleiman-Weiner, Lewis, Liang, Muchovej, Vientós, Young, and Evans]{graziul_does_2023}
Graziul, C., Belikov, A., Chattopadyay, I., Chen, Z., Fang, H., Girdhar, A., Jia, X., Krafft, P.~M., Kleiman-Weiner, M., Lewis, C., Liang, C., Muchovej, J., Vientós, A., Young, M., and Evans, J.
\newblock Does big data serve policy? {Not} without context. {An} experiment with in silico social science.
\newblock \emph{Computational and Mathematical Organization Theory}, 29\penalty0 (1):\penalty0 188--219, March 2023.
\newblock ISSN 1572-9346.
\newblock \doi{10.1007/s10588-022-09362-3}.
\newblock URL \url{https://doi.org/10.1007/s10588-022-09362-3}.

\bibitem[Green(2021)]{green_data_2021}
Green, B.
\newblock Data {Science} as {Political} {Action}: {Grounding} {Data} {Science} in a {Politics} of {Justice}.
\newblock \emph{Journal of Social Computing}, 2\penalty0 (3):\penalty0 249--265, September 2021.
\newblock ISSN 2688-5255.
\newblock \doi{10.23919/JSC.2021.0029}.
\newblock URL \url{https://ieeexplore.ieee.org/abstract/document/9684742}.

\bibitem[Grossman(2023)]{venturebeat2023agi}
Grossman, G.
\newblock {AGI} is coming faster than we think: We must get ready now.
\newblock \emph{VentureBeat}, 2023.
\newblock URL \url{https://venturebeat.com/ai/agi-is-coming-faster-than-we-think-we-must-get-ready-now/}.
\newblock Accessed: Jan 17, 2025.

\bibitem[Gruetzemacher \& Whittlestone(2022)Gruetzemacher and Whittlestone]{gruetzemacher_transformative_2022}
Gruetzemacher, R. and Whittlestone, J.
\newblock The transformative potential of artificial intelligence.
\newblock \emph{Futures}, 135:\penalty0 102884, January 2022.
\newblock ISSN 00163287.
\newblock \doi{10.1016/j.futures.2021.102884}.
\newblock URL \url{https://linkinghub.elsevier.com/retrieve/pii/S0016328721001932}.

\bibitem[Gubrud(1997)]{gubrud_nanotechnology_1997}
Gubrud, M.~A.
\newblock Nanotechnology and {International} {Security}.
\newblock In \emph{Fifth Foresight Conference on Molecular Nanotechnology}, volume~1, 1997.
\newblock URL \url{https://web.archive.org/web/20110529215447/http://www.foresight.org/Conferences/MNT05/Papers/Gubrud/}.

\bibitem[Guest \& Martin(2024)Guest and Martin]{guest_metatheory_2024}
Guest, O. and Martin, A.~E.
\newblock A {Metatheory} of {Classical} and {Modern} {Connectionism}, October 2024.
\newblock URL \url{https://osf.io/eaf2z}.

\bibitem[Gurnee \& Tegmark(2024)Gurnee and Tegmark]{gurnee2024languagemodelsrepresentspace}
Gurnee, W. and Tegmark, M.
\newblock Language models represent space and time.
\newblock In \emph{The Twelfth International Conference on Learning Representations}, 2024.
\newblock URL \url{https://openreview.net/forum?id=jE8xbmvFin}.

\bibitem[Haigh(2024)]{haigh-bust}
Haigh, T.
\newblock How the {AI} boom went bust.
\newblock \emph{Commun. ACM}, 67\penalty0 (2):\penalty0 22–26, January 2024.
\newblock ISSN 0001-0782.
\newblock \doi{10.1145/3634901}.
\newblock URL \url{https://doi.org/10.1145/3634901}.

\bibitem[Hanneke \& Kpotufe(2022)Hanneke and Kpotufe]{hanneke2022no}
Hanneke, S. and Kpotufe, S.
\newblock A no-free-lunch theorem for multitask learning.
\newblock \emph{The Annals of Statistics}, 50\penalty0 (6):\penalty0 3119--3143, 2022.

\bibitem[Hao(2023)]{KarenHaoPanel_2023}
Hao, K.
\newblock The democracy summit 2023, 2023.
\newblock URL \url{https://www.youtube.com/live/0fkGiZ0WqRc?si=NZ9hdvOQLcHNyC4Q&t=28498}.
\newblock [Panel video online; accessed 17-January-2025].

\bibitem[Harrigian et~al.(2023)Harrigian, Zirikly, Chee, Ahmad, Links, Saha, Beach, and Dredze]{harrigian_characterization_2023}
Harrigian, K., Zirikly, A., Chee, B., Ahmad, A., Links, A., Saha, S., Beach, M.~C., and Dredze, M.
\newblock Characterization of {Stigmatizing} {Language} in {Medical} {Records}.
\newblock In Rogers, A., Boyd-Graber, J., and Okazaki, N. (eds.), \emph{Proceedings of the 61st {Annual} {Meeting} of the {Association} for {Computational} {Linguistics} ({Volume} 2: {Short} {Papers})}, pp.\  312--329, Toronto, Canada, July 2023. Association for Computational Linguistics.
\newblock \doi{10.18653/v1/2023.acl-short.28}.
\newblock URL \url{https://aclanthology.org/2023.acl-short.28/}.

\bibitem[Henshall(2024)]{henshall_when_2024}
Henshall, W.
\newblock When {Might} {AI} {Outsmart} {Us}? {It} {Depends} {Who} {You} {Ask}.
\newblock \emph{Time}, January 2024.
\newblock URL \url{https://time.com/6556168/when-ai-outsmart-humans/}.

\bibitem[Hernández-Orallo \& Seán Ó~hÉigeartaigh(2018)Hernández-Orallo and Seán Ó~hÉigeartaigh]{hernandez2018paradigms}
Hernández-Orallo, J. and Seán Ó~hÉigeartaigh, S.
\newblock Paradigms of artificial general intelligence and their associated risks.
\newblock \emph{Centre for the Study of Existential Risk, University of Cambridge, UK}, 2018.

\bibitem[Hernández-Orallo et~al.(2014)Hernández-Orallo, Dowe, and Hernández-Lloreda]{HERNANDEZORALLO201450}
Hernández-Orallo, J., Dowe, D.~L., and Hernández-Lloreda, M.
\newblock Universal psychometrics: Measuring cognitive abilities in the machine kingdom.
\newblock \emph{Cognitive Systems Research}, 27:\penalty0 50--74, 2014.
\newblock ISSN 1389-0417.
\newblock \doi{https://doi.org/10.1016/j.cogsys.2013.06.001}.
\newblock URL \url{https://www.sciencedirect.com/science/article/pii/S1389041713000338}.

\bibitem[Hernández-Orallo et~al.(2021)Hernández-Orallo, Loe, Cheke, Martínez-Plumed, and Ó~hÉigeartaigh]{hernandez-orallo_general_2021}
Hernández-Orallo, J., Loe, B.~S., Cheke, L., Martínez-Plumed, F., and Ó~hÉigeartaigh, S.
\newblock General intelligence disentangled via a generality metric for natural and artificial intelligence.
\newblock \emph{Scientific Reports}, 11\penalty0 (1):\penalty0 22822, November 2021.
\newblock ISSN 2045-2322.
\newblock \doi{10.1038/s41598-021-01997-7}.
\newblock URL \url{https://www.nature.com/articles/s41598-021-01997-7}.

\bibitem[Herrmann et~al.(2024)Herrmann, Lange, Eggensperger, Casalicchio, Wever, Feurer, Rügamer, Hüllermeier, Boulesteix, and Bischl]{herrmann_position_2024}
Herrmann, M., Lange, F. J.~D., Eggensperger, K., Casalicchio, G., Wever, M., Feurer, M., Rügamer, D., Hüllermeier, E., Boulesteix, A.-L., and Bischl, B.
\newblock Position: {Why} {We} {Must} {Rethink} {Empirical} {Research} in {Machine} {Learning}.
\newblock In \emph{Proceedings of the 41st {International} {Conference} on {Machine} {Learning}}, pp.\  18228--18247. PMLR, July 2024.
\newblock URL \url{https://proceedings.mlr.press/v235/herrmann24b.html}.
\newblock ISSN: 2640-3498.

\bibitem[Hewlett et~al.(2013)Hewlett, Marshall, and Sherbin]{hewlett_how_2013}
Hewlett, S.~A., Marshall, M., and Sherbin, L.
\newblock How {Diversity} {Can} {Drive} {Innovation}.
\newblock \emph{Harvard Business Review}, 91\penalty0 (12), December 2013.
\newblock ISSN 0017-8012.

\bibitem[Hicks et~al.(2024)Hicks, Humphries, and Slater]{hicks_chatgpt_2024}
Hicks, M.~T., Humphries, J., and Slater, J.
\newblock {ChatGPT} is bullshit.
\newblock \emph{Ethics and Information Technology}, 26\penalty0 (2):\penalty0 38, June 2024.
\newblock ISSN 1572-8439.
\newblock \doi{10.1007/s10676-024-09775-5}.
\newblock URL \url{https://doi.org/10.1007/s10676-024-09775-5}.

\bibitem[Holland(2025)]{reuters2025trumpai}
Holland, S.
\newblock Trump to announce private sector {AI} infrastructure investment, {CBS} reports.
\newblock \emph{Reuters}, January 2025.
\newblock URL \url{https://www.reuters.com/technology/artificial-intelligence/trump-announce-private-sector-ai-infrastructure-investment-cbs-reports-2025-01-21/}.

\bibitem[Hong \& Page(2004)Hong and Page]{hong_groups_2004}
Hong, L. and Page, S.~E.
\newblock Groups of diverse problem solvers can outperform groups of high-ability problem solvers.
\newblock \emph{Proceedings of the National Academy of Sciences}, 101\penalty0 (46):\penalty0 16385--16389, November 2004.
\newblock \doi{10.1073/pnas.0403723101}.
\newblock URL \url{https://www.pnas.org/doi/full/10.1073/pnas.0403723101}.

\bibitem[Hooker(2021)]{hooker_hardware_2021}
Hooker, S.
\newblock The hardware lottery.
\newblock \emph{Commun. ACM}, 64\penalty0 (12):\penalty0 58--65, November 2021.
\newblock ISSN 0001-0782.
\newblock \doi{10.1145/3467017}.
\newblock URL \url{https://doi.org/10.1145/3467017}.

\bibitem[Hooker(2024)]{hooker_diminishing_2024}
Hooker, S.
\newblock On the diminishing returns to scaling.
\newblock [Online - Accessed 2024-01-12], Nov 2024.
\newblock URL \url{https://drive.google.com/file/d/1yeW429nx_FXaK_RgqDv89wH4Gh5flIRG/view}.

\bibitem[Hullman et~al.(2022)Hullman, Kapoor, Nanayakkara, Gelman, and Narayanan]{hullman_worst}
Hullman, J., Kapoor, S., Nanayakkara, P., Gelman, A., and Narayanan, A.
\newblock The worst of both worlds: A comparative analysis of errors in learning from data in psychology and machine learning.
\newblock In \emph{Proceedings of the 2022 AAAI/ACM Conference on AI, Ethics, and Society}, AIES '22, pp.\  335–348, New York, NY, USA, 2022. Association for Computing Machinery.
\newblock ISBN 9781450392471.
\newblock \doi{10.1145/3514094.3534196}.
\newblock URL \url{https://doi.org/10.1145/3514094.3534196}.

\bibitem[Hutchinson et~al.(2022)Hutchinson, Rostamzadeh, Greer, Heller, and Prabhakaran]{hutchinson_evalgaps}
Hutchinson, B., Rostamzadeh, N., Greer, C., Heller, K., and Prabhakaran, V.
\newblock Evaluation gaps in machine learning practice.
\newblock In \emph{Proceedings of the 2022 ACM Conference on Fairness, Accountability, and Transparency}, FAccT '22, pp.\  1859–1876, New York, NY, USA, 2022. Association for Computing Machinery.
\newblock ISBN 9781450393522.
\newblock \doi{10.1145/3531146.3533233}.
\newblock URL \url{https://doi.org/10.1145/3531146.3533233}.

\bibitem[IBM(2023)]{ibm2023agi}
IBM.
\newblock Getting ready for artificial general intelligence with examples, 2023.
\newblock URL \url{https://www.ibm.com/think/topics/artificial-general-intelligence-examples}.
\newblock Accessed: Jan 17, 2025.

\bibitem[Jacobs \& Wallach(2021)Jacobs and Wallach]{jacobs_measurement_2021}
Jacobs, A.~Z. and Wallach, H.
\newblock Measurement and {Fairness}.
\newblock In \emph{Proceedings of the 2021 {ACM} {Conference} on {Fairness}, {Accountability}, and {Transparency}}, pp.\  375--385, Virtual Event Canada, March 2021. ACM.
\newblock ISBN 978-1-4503-8309-7.
\newblock \doi{10.1145/3442188.3445901}.
\newblock URL \url{https://dl.acm.org/doi/10.1145/3442188.3445901}.

\bibitem[Jain et~al.(2024)Jain, Suriyakumar, Creel, and Wilson]{jain_algorithmic_2024}
Jain, S., Suriyakumar, V., Creel, K., and Wilson, A.
\newblock Algorithmic {Pluralism}: {A} {Structural} {Approach} {To} {Equal} {Opportunity}.
\newblock In \emph{Proceedings of the 2024 {ACM} {Conference} on {Fairness}, {Accountability}, and {Transparency}}, {FAccT} '24, pp.\  197--206, New York, NY, USA, June 2024. Association for Computing Machinery.
\newblock ISBN 9798400704505.
\newblock \doi{10.1145/3630106.3658899}.
\newblock URL \url{https://dl.acm.org/doi/10.1145/3630106.3658899}.

\bibitem[Jones(2020)]{jones2020law}
Jones, C.
\newblock Law enforcement use of facial recognition: bias, disparate impacts on people of color, and the need for federal legislation.
\newblock \emph{NCJL \& Tech.}, 22:\penalty0 777, 2020.

\bibitem[Jones(2025)]{jones_how_2025}
Jones, N.
\newblock How should we test {AI} for human-level intelligence? {OpenAI}’s o3 electrifies quest.
\newblock \emph{Nature}, 637\penalty0 (8047):\penalty0 774--775, January 2025.
\newblock ISSN 1476-4687.
\newblock \doi{10.1038/d41586-025-00110-6}.
\newblock URL \url{https://www.nature.com/articles/d41586-025-00110-6}.

\bibitem[Kaack et~al.(2022)Kaack, Donti, Strubell, Kamiya, Creutzig, and Rolnick]{kaack2022aligning}
Kaack, L.~H., Donti, P.~L., Strubell, E., Kamiya, G., Creutzig, F., and Rolnick, D.
\newblock Aligning artificial intelligence with climate change mitigation.
\newblock \emph{Nature Climate Change}, 12\penalty0 (6):\penalty0 518--527, 2022.

\bibitem[Kelly(2024)]{cnn2024muskai}
Kelly, S.~M.
\newblock Elon {Musk} says {AI} will take your job, and 'no one is going to need to work'.
\newblock \emph{CNN}, May 2024.
\newblock URL \url{https://www.cnn.com/2024/05/23/tech/elon-musk-ai-your-job/index.html}.
\newblock Accessed: Janury 19, 2025.

\bibitem[Kerner(2023)]{musk-xai}
Kerner, S.~M.
\newblock {Elon Musk} reveals {xAI} efforts, predicts full {AGI} by 2029, 2023.
\newblock URL \url{https://venturebeat.com/ai/elon-musk-reveals-xai-efforts-predicts-full-agi-by-2029/}.
\newblock [Online; accessed 19-May-2025].

\bibitem[Kierans et~al.(2025)Kierans, Ghosh, Hazan, and Dori-Hacohen]{kierans2024quantifyingmisalignmentagentssociotechnical}
Kierans, A., Ghosh, A., Hazan, H., and Dori-Hacohen, S.
\newblock Quantifying misalignment between agents: Towards a sociotechnical understanding of alignment.
\newblock \emph{Proceedings of the AAAI Conference on Artificial Intelligence}, March 2025.
\newblock URL \url{https://arxiv.org/abs/2406.04231}.

\bibitem[Klein(2025)]{klein_opinion_2025}
Klein, E.
\newblock Opinion {\textbar} {The} {Government} {Knows} {A}.{G}.{I}. {Is} {Coming}.
\newblock \emph{The New York Times}, March 2025.
\newblock ISSN 0362-4331.
\newblock URL \url{https://www.nytimes.com/2025/03/04/opinion/ezra-klein-podcast-ben-buchanan.html}.

\bibitem[Kleinberg \& Raghavan(2021)Kleinberg and Raghavan]{kleinberg_algorithmic2021}
Kleinberg, J. and Raghavan, M.
\newblock Algorithmic monoculture and social welfare.
\newblock \emph{Proceedings of the National Academy of Sciences}, 118\penalty0 (22):\penalty0 e2018340118, 2021.
\newblock \doi{10.1073/pnas.2018340118}.
\newblock URL \url{https://www.pnas.org/doi/abs/10.1073/pnas.2018340118}.

\bibitem[Knorr~Cetina(1999)]{knorr_cetina_epistemic_1999}
Knorr~Cetina, K.
\newblock \emph{Epistemic {Cultures}: {How} the {Sciences} {Make} {Knowledge}}.
\newblock Harvard University Press, May 1999.
\newblock ISBN 978-0-674-03968-1.

\bibitem[Knorr~Cetina(2007)]{knorr_cetina_culture_2007}
Knorr~Cetina, K.
\newblock Culture in global knowledge societies: knowledge cultures and epistemic cultures.
\newblock \emph{Interdisciplinary Science Reviews}, 32\penalty0 (4):\penalty0 361--375, December 2007.
\newblock ISSN 0308-0188.
\newblock \doi{10.1179/030801807X163571}.
\newblock URL \url{https://journals.sagepub.com/doi/abs/10.1179/030801807X163571}.

\bibitem[Kwon \& Porter(2025)Kwon and Porter]{kwon_use_2025}
Kwon, S. and Porter, A.~L.
\newblock Use of exclusive data for corporate research on machine learning and artificial intelligence: {Implications} for innovation and competition policy.
\newblock \emph{Technology in Society}, 81:\penalty0 102820, June 2025.
\newblock ISSN 0160-791X.
\newblock \doi{10.1016/j.techsoc.2025.102820}.
\newblock URL \url{https://www.sciencedirect.com/science/article/pii/S0160791X25000107}.

\bibitem[LaForge(2024)]{laforge_dangers_2024}
LaForge, G.
\newblock The {Dangers} of {Imposing} {Global} {North} {Approaches} to {AI} {Governance} on the {Global} {South} {\textbar} {TechPolicy}.{Press}, September 2024.
\newblock URL \url{https://techpolicy.press/the-dangers-of-imposing-global-north-approaches-to-ai-governance-on-the-global-south/}.

\bibitem[Lazar(2022)]{lazar_power_2022}
Lazar, S.
\newblock Power and {AI}: {Nature} and {Justification}.
\newblock In Bullock, J., Chen, Y.-C., Himmelreich, J., Hudson, V.~M., Korinek, A., Young, M., and Zhang, B. (eds.), \emph{The {Oxford} {Handbook} of {AI} {Governance}}. Oxford University Press, May 2022.
\newblock ISBN 978-0-19-757932-9.
\newblock \doi{10.1093/oxfordhb/9780197579329.013.12}.
\newblock URL \url{https://oxfordhandbooks.com/view/10.1093/oxfordhb/9780197579329.001.0001/oxfordhb-9780197579329-e-12}.

\bibitem[Lazar \& Nelson(2023)Lazar and Nelson]{lazar_ai_2023}
Lazar, S. and Nelson, A.
\newblock {AI} safety on whose terms?
\newblock \emph{Science}, 381\penalty0 (6654):\penalty0 138--138, July 2023.
\newblock \doi{10.1126/science.adi8982}.
\newblock URL \url{https://www.science.org/doi/10.1126/science.adi8982}.

\bibitem[Legg \& Hutter(2007)Legg and Hutter]{legg_universal_2007}
Legg, S. and Hutter, M.
\newblock Universal {Intelligence}: {A} {Definition} of {Machine} {Intelligence}.
\newblock \emph{Minds and Machines}, 17\penalty0 (4):\penalty0 391--444, December 2007.
\newblock ISSN 1572-8641.
\newblock \doi{10.1007/s11023-007-9079-x}.
\newblock URL \url{https://doi.org/10.1007/s11023-007-9079-x}.

\bibitem[Leong \& Linzen(2024)Leong and Linzen]{leong_testing_2024}
Leong, C. S.-Y. and Linzen, T.
\newblock Testing learning hypotheses using neural networks by manipulating learning data, July 2024.
\newblock URL \url{http://arxiv.org/abs/2407.04593}.
\newblock arXiv:2407.04593 [cs].

\bibitem[Liao et~al.(2021)Liao, Taori, Raji, and Schmidt]{liao_are_2021}
Liao, T., Taori, R., Raji, D., and Schmidt, L.
\newblock Are {We} {Learning} {Yet}? {A} {Meta} {Review} of {Evaluation} {Failures} {Across} {Machine} {Learning}.
\newblock \emph{Proceedings of the Neural Information Processing Systems Track on Datasets and Benchmarks}, 1, 2021.
\newblock URL \url{https://datasets-benchmarks-proceedings.neurips.cc/paper/2021/file/757b505cfd34c64c85ca5b5690ee5293-Paper-round2.pdf}.

\bibitem[Liebowitz \& Margolis(1995)Liebowitz and Margolis]{liebowitz_path_1995}
Liebowitz, S.~J. and Margolis, S.~E.
\newblock Path {Dependence}, {Lock}-in, and {History}.
\newblock \emph{Journal of Law, Economics, \& Organization}, 11\penalty0 (1):\penalty0 205--226, 1995.
\newblock ISSN 87566222, 14657341.
\newblock URL \url{http://www.jstor.org/stable/765077}.

\bibitem[Lin et~al.(2020)Lin, Yu, Zhou, Zhou, and Shi]{lin_how_2020}
Lin, J., Yu, Y., Zhou, Y., Zhou, Z., and Shi, X.
\newblock How many preprints have actually been printed and why: a case study of computer science preprints on {arXiv}.
\newblock \emph{Scientometrics}, 124\penalty0 (1):\penalty0 555--574, July 2020.
\newblock ISSN 1588-2861.
\newblock \doi{10.1007/s11192-020-03430-8}.
\newblock URL \url{https://doi.org/10.1007/s11192-020-03430-8}.

\bibitem[Luccioni et~al.(2024)Luccioni, Gamazaychikov, Hooker, Pierrard, Strubell, Jernite, and Wu]{luccioni2024light}
Luccioni, S., Gamazaychikov, B., Hooker, S., Pierrard, R., Strubell, E., Jernite, Y., and Wu, C.-J.
\newblock Light bulbs have energy ratings---so why can't {AI} chatbots?
\newblock \emph{Nature}, 632\penalty0 (8026):\penalty0 736--738, 2024.

\bibitem[Marcus(2022)]{Marcus_2022}
Marcus, G.
\newblock Dear {Elon Musk}, here are five things you might want to consider about {AGI}, 2022.
\newblock URL \url{https://garymarcus.substack.com/p/dear-elon-musk-here-are-five-things}.
\newblock [Online; accessed 24-January-2024].

\bibitem[Mathur et~al.(2022)Mathur, Lustig, and Kaziunas]{mathur_disordering_2022}
Mathur, V., Lustig, C., and Kaziunas, E.
\newblock Disordering {Datasets}: {Sociotechnical} {Misalignments} in {AI}-{Mediated} {Behavioral} {Health}.
\newblock \emph{Proceedings of the ACM on Human-Computer Interaction}, 6\penalty0 (CSCW2):\penalty0 1--33, November 2022.
\newblock ISSN 2573-0142.
\newblock \doi{10.1145/3555141}.
\newblock URL \url{https://dl.acm.org/doi/10.1145/3555141}.

\bibitem[Maymin(2023)]{forbes-hinton}
Maymin, P.
\newblock Artificial general intelligence ({AGI}) is one prompt away, 2023.
\newblock URL \url{https://www.forbes.com/sites/philipmaymin/2023/10/13/artificial-general-intelligence-agi-is-one-prompt-away/}.
\newblock [Online; accessed 19-May-2025].

\bibitem[McCarthy \& Hayes(1981)McCarthy and Hayes]{mccarthy1981some}
McCarthy, J. and Hayes, P.~J.
\newblock Some philosophical problems from the standpoint of artificial intelligence.
\newblock In \emph{Readings in artificial intelligence}, pp.\  431--450. Elsevier, 1981.

\bibitem[McCarthy et~al.(1955)McCarthy, Minsky, Rochester, and Shannon]{mccarthy_proposal_1955}
McCarthy, J., Minsky, M.~L., Rochester, N., and Shannon, C.
\newblock A {Proposal} for the {Dartmouth} {Summer} {Research} {Project} on {Artificial} {Intelligence}, August 1955.
\newblock URL \url{http://jmc.stanford.edu/articles/dartmouth/dartmouth.pdf}.

\bibitem[Midgley(2000)]{midgley_methodological_2000}
Midgley, G.
\newblock Methodological {Pluralism}.
\newblock In Minati, G., Giuliani, A., and Bich, L. (eds.), \emph{Systemic {Intervention}: {Philosophy}, {Methodology}, and {Practice}}, Contemporary {Systems} {Thinking}, pp.\  171--216. Springer US, Boston, MA, 2000.
\newblock ISBN 978-1-4615-4201-8.
\newblock \doi{10.1007/978-1-4615-4201-8_9}.
\newblock URL \url{https://doi.org/10.1007/978-1-4615-4201-8_9}.

\bibitem[Mikesell et~al.(2013)Mikesell, Bromley, and Khodyakov]{mikesell_ethical_2013}
Mikesell, L., Bromley, E., and Khodyakov, D.
\newblock Ethical {Community}-{Engaged} {Research}: {A} {Literature} {Review}.
\newblock \emph{American Journal of Public Health}, 103\penalty0 (12):\penalty0 e7--e14, December 2013.
\newblock ISSN 0090-0036.
\newblock \doi{10.2105/AJPH.2013.301605}.
\newblock URL \url{https://ajph.aphapublications.org/doi/full/10.2105/AJPH.2013.301605}.

\bibitem[Mitchell(2024)]{mitchell_debates_2024}
Mitchell, M.
\newblock Debates on the nature of artificial general intelligence.
\newblock \emph{Science}, 383\penalty0 (6689):\penalty0 eado7069, March 2024.
\newblock ISSN 0036-8075, 1095-9203.
\newblock \doi{10.1126/science.ado7069}.
\newblock URL \url{https://www.science.org/doi/10.1126/science.ado7069}.

\bibitem[Morris et~al.(2024)Morris, Sohl-Dickstein, Fiedel, Warkentin, Dafoe, Faust, Farabet, and Legg]{morris_position_2024}
Morris, M.~R., Sohl-Dickstein, J., Fiedel, N., Warkentin, T., Dafoe, A., Faust, A., Farabet, C., and Legg, S.
\newblock Position: levels of {AGI} for operationalizing progress on the path to {AGI}.
\newblock In \emph{Proceedings of the 41st {International} {Conference} on {Machine} {Learning}}, volume 235 of \emph{{ICML}'24}, pp.\  36308--36321, Vienna, Austria, July 2024. JMLR.org.

\bibitem[Mueller(2024)]{mueller2024myth}
Mueller, M.
\newblock The myth of {AGI}.
\newblock \emph{Internet Governance Project}, 2024.
\newblock URL \url{https://www.internetgovernance.org/wp-content/uploads/MythofAGI.pdf}.

\bibitem[Muldoon(2013)]{muldoon_diversity_2013}
Muldoon, R.
\newblock Diversity and the {Division} of {Cognitive} {Labor}.
\newblock \emph{Philosophy Compass}, 8\penalty0 (2):\penalty0 117--125, 2013.
\newblock ISSN 1747-9991.
\newblock \doi{10.1111/phc3.12000}.
\newblock URL \url{https://onlinelibrary.wiley.com/doi/abs/10.1111/phc3.12000}.

\bibitem[Mulligan et~al.(2016)Mulligan, Koopman, and Doty]{mulligan_privacy_2016}
Mulligan, D.~K., Koopman, C., and Doty, N.
\newblock Privacy is an essentially contested concept: a multi-dimensional analytic for mapping privacy.
\newblock \emph{Philosophical Transactions of the Royal Society A: Mathematical, Physical and Engineering Sciences}, 374\penalty0 (2083):\penalty0 20160118, December 2016.
\newblock \doi{10.1098/rsta.2016.0118}.
\newblock URL \url{https://royalsocietypublishing.org/doi/10.1098/rsta.2016.0118}.
\newblock Publisher: Royal Society.

\bibitem[Narayanan \& Kapoor(2024)Narayanan and Kapoor]{narayanan_ai_2024}
Narayanan, A. and Kapoor, S.
\newblock \emph{{AI} {Snake} {Oil}: {What} {Artificial} {Intelligence} {Can} {Do}, {What} {It} {Can}’t, and {How} to {Tell} the {Difference}}.
\newblock Princeton University Press, September 2024.
\newblock ISBN 978-0-691-24964-3.
\newblock \doi{10.1515/9780691249643}.
\newblock URL \url{https://www.degruyter.com/document/doi/10.1515/9780691249643/html}.

\bibitem[Newell \& Ernst(1965)Newell and Ernst]{newell1965search}
Newell, A. and Ernst, G.
\newblock The search for generality.
\newblock In \emph{Proc. IFIP Congress}, volume~65, pp.\  17--24, 1965.

\bibitem[Nilsson(2005)]{nilsson_human-level_2005}
Nilsson, N.~J.
\newblock Human-{Level} {Artificial} {Intelligence}? {Be} {Serious}!
\newblock \emph{AI Magazine}, 26\penalty0 (4):\penalty0 68--68, December 2005.
\newblock ISSN 2371-9621.
\newblock \doi{10.1609/aimag.v26i4.1850}.
\newblock URL \url{https://ojs.aaai.org/aimagazine/index.php/aimagazine/article/view/1850}.

\bibitem[Obermeyer et~al.(2019)Obermeyer, Powers, Vogeli, and Mullainathan]{obermeyer_dissecting_2019}
Obermeyer, Z., Powers, B., Vogeli, C., and Mullainathan, S.
\newblock Dissecting racial bias in an algorithm used to manage the health of populations.
\newblock \emph{Science}, 366\penalty0 (6464):\penalty0 447--453, October 2019.
\newblock \doi{10.1126/science.aax2342}.
\newblock URL \url{https://www.science.org/doi/full/10.1126/science.aax2342}.
\newblock Publisher: American Association for the Advancement of Science.

\bibitem[OpenAI(2018)]{openai_2018}
OpenAI.
\newblock {OpenAI} {Charter}.
\newblock Technical report, OpenAI, April 2018.
\newblock URL \url{https://openai.com/charter}.

\bibitem[OpenAI(2025{\natexlab{a}})]{OpenAIAbout}
OpenAI.
\newblock About, 2025{\natexlab{a}}.
\newblock URL \url{https://openai.com/about/}.
\newblock [Online; accessed 19-May-2025].

\bibitem[OpenAI(2025{\natexlab{b}})]{OpenAIPlanning}
OpenAI.
\newblock Planning for {AGI} and beyond, 2025{\natexlab{b}}.
\newblock URL \url{https://openai.com/index/planning-for-agi-and-beyond/}.
\newblock [Online; accessed 19-May-2025].

\bibitem[OpenAI(2025{\natexlab{c}})]{OpenAISecurity}
OpenAI.
\newblock Security on the path to {AGI}, 2025{\natexlab{c}}.
\newblock URL \url{https://openai.com/index/security-on-the-path-to-agi/}.
\newblock [Online; accessed 19-May-2025].

\bibitem[OpenAI(2025{\natexlab{d}})]{OpenAIStructure}
OpenAI.
\newblock Our structure, 2025{\natexlab{d}}.
\newblock URL \url{https://openai.com/our-structure/}.
\newblock [Online; accessed 17-January-2025].

\bibitem[OpenAI et~al.(2024)OpenAI, Achiam, Adler, Agarwal, Ahmad, Akkaya, Aleman, Almeida, Altenschmidt, Altman, Anadkat, Avila, Babuschkin, Balaji, Balcom, Baltescu, Bao, Bavarian, Belgum, Bello, Berdine, Bernadett-Shapiro, Berner, Bogdonoff, Boiko, Boyd, Brakman, Brockman, Brooks, Brundage, Button, Cai, Campbell, Cann, Carey, Carlson, Carmichael, Chan, Chang, Chantzis, Chen, Chen, Chen, Chen, Chen, Chess, Cho, Chu, Chung, Cummings, Currier, Dai, Decareaux, Degry, Deutsch, Deville, Dhar, Dohan, Dowling, Dunning, Ecoffet, Eleti, Eloundou, Farhi, Fedus, Felix, Fishman, Forte, Fulford, Gao, Georges, Gibson, Goel, Gogineni, Goh, Gontijo-Lopes, Gordon, Grafstein, Gray, Greene, Gross, Gu, Guo, Hallacy, Han, Harris, He, Heaton, Heidecke, Hesse, Hickey, Hickey, Hoeschele, Houghton, Hsu, Hu, Hu, Huizinga, Jain, Jain, Jang, Jiang, Jiang, Jin, Jin, Jomoto, Jonn, Jun, Kaftan, Łukasz Kaiser, Kamali, Kanitscheider, Keskar, Khan, Kilpatrick, Kim, Kim, Kim, Kirchner, Kiros, Knight, Kokotajlo, Łukasz Kondraciuk, Kondrich,
  Konstantinidis, Kosic, Krueger, Kuo, Lampe, Lan, Lee, Leike, Leung, Levy, Li, Lim, Lin, Lin, Litwin, Lopez, Lowe, Lue, Makanju, Malfacini, Manning, Markov, Markovski, Martin, Mayer, Mayne, McGrew, McKinney, McLeavey, McMillan, McNeil, Medina, Mehta, Menick, Metz, Mishchenko, Mishkin, Monaco, Morikawa, Mossing, Mu, Murati, Murk, Mély, Nair, Nakano, Nayak, Neelakantan, Ngo, Noh, Ouyang, O'Keefe, Pachocki, Paino, Palermo, Pantuliano, Parascandolo, Parish, Parparita, Passos, Pavlov, Peng, Perelman, de~Avila Belbute~Peres, Petrov, de~Oliveira~Pinto, Michael, Pokorny, Pokrass, Pong, Powell, Power, Power, Proehl, Puri, Radford, Rae, Ramesh, Raymond, Real, Rimbach, Ross, Rotsted, Roussez, Ryder, Saltarelli, Sanders, Santurkar, Sastry, Schmidt, Schnurr, Schulman, Selsam, Sheppard, Sherbakov, Shieh, Shoker, Shyam, Sidor, Sigler, Simens, Sitkin, Slama, Sohl, Sokolowsky, Song, Staudacher, Such, Summers, Sutskever, Tang, Tezak, Thompson, Tillet, Tootoonchian, Tseng, Tuggle, Turley, Tworek, Uribe, Vallone, Vijayvergiya,
  Voss, Wainwright, Wang, Wang, Wang, Ward, Wei, Weinmann, Welihinda, Welinder, Weng, Weng, Wiethoff, Willner, Winter, Wolrich, Wong, Workman, Wu, Wu, Wu, Xiao, Xu, Yoo, Yu, Yuan, Zaremba, Zellers, Zhang, Zhang, Zhao, Zheng, Zhuang, Zhuk, and Zoph]{openai2024gpt4technicalreport}
OpenAI, Achiam, J., Adler, S., Agarwal, S., Ahmad, L., Akkaya, I., Aleman, F.~L., Almeida, D., Altenschmidt, J., Altman, S., Anadkat, S., Avila, R., Babuschkin, I., Balaji, S., Balcom, V., Baltescu, P., Bao, H., Bavarian, M., Belgum, J., Bello, I., Berdine, J., Bernadett-Shapiro, G., Berner, C., Bogdonoff, L., Boiko, O., Boyd, M., Brakman, A.-L., Brockman, G., Brooks, T., Brundage, M., Button, K., Cai, T., Campbell, R., Cann, A., Carey, B., Carlson, C., Carmichael, R., Chan, B., Chang, C., Chantzis, F., Chen, D., Chen, S., Chen, R., Chen, J., Chen, M., Chess, B., Cho, C., Chu, C., Chung, H.~W., Cummings, D., Currier, J., Dai, Y., Decareaux, C., Degry, T., Deutsch, N., Deville, D., Dhar, A., Dohan, D., Dowling, S., Dunning, S., Ecoffet, A., Eleti, A., Eloundou, T., Farhi, D., Fedus, L., Felix, N., Fishman, S.~P., Forte, J., Fulford, I., Gao, L., Georges, E., Gibson, C., Goel, V., Gogineni, T., Goh, G., Gontijo-Lopes, R., Gordon, J., Grafstein, M., Gray, S., Greene, R., Gross, J., Gu, S.~S., Guo, Y., Hallacy,
  C., Han, J., Harris, J., He, Y., Heaton, M., Heidecke, J., Hesse, C., Hickey, A., Hickey, W., Hoeschele, P., Houghton, B., Hsu, K., Hu, S., Hu, X., Huizinga, J., Jain, S., Jain, S., Jang, J., Jiang, A., Jiang, R., Jin, H., Jin, D., Jomoto, S., Jonn, B., Jun, H., Kaftan, T., Łukasz Kaiser, Kamali, A., Kanitscheider, I., Keskar, N.~S., Khan, T., Kilpatrick, L., Kim, J.~W., Kim, C., Kim, Y., Kirchner, J.~H., Kiros, J., Knight, M., Kokotajlo, D., Łukasz Kondraciuk, Kondrich, A., Konstantinidis, A., Kosic, K., Krueger, G., Kuo, V., Lampe, M., Lan, I., Lee, T., Leike, J., Leung, J., Levy, D., Li, C.~M., Lim, R., Lin, M., Lin, S., Litwin, M., Lopez, T., Lowe, R., Lue, P., Makanju, A., Malfacini, K., Manning, S., Markov, T., Markovski, Y., Martin, B., Mayer, K., Mayne, A., McGrew, B., McKinney, S.~M., McLeavey, C., McMillan, P., McNeil, J., Medina, D., Mehta, A., Menick, J., Metz, L., Mishchenko, A., Mishkin, P., Monaco, V., Morikawa, E., Mossing, D., Mu, T., Murati, M., Murk, O., Mély, D., Nair, A., Nakano, R.,
  Nayak, R., Neelakantan, A., Ngo, R., Noh, H., Ouyang, L., O'Keefe, C., Pachocki, J., Paino, A., Palermo, J., Pantuliano, A., Parascandolo, G., Parish, J., Parparita, E., Passos, A., Pavlov, M., Peng, A., Perelman, A., de~Avila Belbute~Peres, F., Petrov, M., de~Oliveira~Pinto, H.~P., Michael, Pokorny, Pokrass, M., Pong, V.~H., Powell, T., Power, A., Power, B., Proehl, E., Puri, R., Radford, A., Rae, J., Ramesh, A., Raymond, C., Real, F., Rimbach, K., Ross, C., Rotsted, B., Roussez, H., Ryder, N., Saltarelli, M., Sanders, T., Santurkar, S., Sastry, G., Schmidt, H., Schnurr, D., Schulman, J., Selsam, D., Sheppard, K., Sherbakov, T., Shieh, J., Shoker, S., Shyam, P., Sidor, S., Sigler, E., Simens, M., Sitkin, J., Slama, K., Sohl, I., Sokolowsky, B., Song, Y., Staudacher, N., Such, F.~P., Summers, N., Sutskever, I., Tang, J., Tezak, N., Thompson, M.~B., Tillet, P., Tootoonchian, A., Tseng, E., Tuggle, P., Turley, N., Tworek, J., Uribe, J. F.~C., Vallone, A., Vijayvergiya, A., Voss, C., Wainwright, C., Wang,
  J.~J., Wang, A., Wang, B., Ward, J., Wei, J., Weinmann, C., Welihinda, A., Welinder, P., Weng, J., Weng, L., Wiethoff, M., Willner, D., Winter, C., Wolrich, S., Wong, H., Workman, L., Wu, S., Wu, J., Wu, M., Xiao, K., Xu, T., Yoo, S., Yu, K., Yuan, Q., Zaremba, W., Zellers, R., Zhang, C., Zhang, M., Zhao, S., Zheng, T., Zhuang, J., Zhuk, W., and Zoph, B.
\newblock Gpt-4 technical report, 2024.
\newblock URL \url{https://arxiv.org/abs/2303.08774}.

\bibitem[Ovadya(2023)]{ovadya_reimagining_2023}
Ovadya, A.
\newblock Reimagining {Democracy} for {AI}.
\newblock \emph{Journal of Democracy}, 34\penalty0 (4):\penalty0 162--170, 2023.
\newblock ISSN 1086-3214.
\newblock \doi{10.1353/jod.2023.a907697}.
\newblock URL \url{https://muse.jhu.edu/pub/1/article/907697}.

\bibitem[Paolo et~al.(2024)Paolo, Gonzalez-Billandon, and K\'{e}gl]{pmlr-v235-paolo24a}
Paolo, G., Gonzalez-Billandon, J., and K\'{e}gl, B.
\newblock Position: A call for embodied {AI}.
\newblock In Salakhutdinov, R., Kolter, Z., Heller, K., Weller, A., Oliver, N., Scarlett, J., and Berkenkamp, F. (eds.), \emph{Proceedings of the 41st International Conference on Machine Learning}, volume 235 of \emph{Proceedings of Machine Learning Research}, pp.\  39493--39508. PMLR, 21--27 Jul 2024.
\newblock URL \url{https://proceedings.mlr.press/v235/paolo24a.html}.

\bibitem[Peacock(2009)]{peacock_path_2009}
Peacock, M.~S.
\newblock Path {Dependence} in the {Production} of {Scientific} {Knowledge}.
\newblock \emph{Social Epistemology}, 23\penalty0 (2):\penalty0 105--124, April 2009.
\newblock ISSN 0269-1728, 1464-5297.
\newblock \doi{10.1080/02691720902962813}.
\newblock URL \url{http://www.tandfonline.com/doi/abs/10.1080/02691720902962813}.

\bibitem[Perrigo(2024)]{time-meta}
Perrigo, B.
\newblock Meta's {AI} chief {Yann LeCun} on {AGI}, open-source, and {AI} risk, 2024.
\newblock URL \url{https://time.com/6694432/yann-lecun-meta-ai-interview/}.
\newblock [Online; accessed 19-May-2025].

\bibitem[Pierre et~al.(2021)Pierre, Crooks, Currie, Paris, and Pasquetto]{pierregetting_2021}
Pierre, J., Crooks, R., Currie, M., Paris, B., and Pasquetto, I.
\newblock Getting {Ourselves} {Together}: {Data}-centered participatory design research \& epistemic burden.
\newblock In \emph{Proceedings of the 2021 {CHI} {Conference} on {Human} {Factors} in {Computing} {Systems}}, {CHI} '21, pp.\  1--11, New York, NY, USA, May 2021. Association for Computing Machinery.
\newblock ISBN 978-1-4503-8096-6.
\newblock \doi{10.1145/3411764.3445103}.
\newblock URL \url{https://dl.acm.org/doi/10.1145/3411764.3445103}.

\bibitem[Pierson et~al.(2025)Pierson, Shanmugam, Movva, Kleinberg, Agrawal, Dredze, Ferryman, Gichoya, Jurafsky, Koh, Levy, Mullainathan, Obermeyer, Suresh, and Vafa]{pierson_using_2025}
Pierson, E., Shanmugam, D., Movva, R., Kleinberg, J., Agrawal, M., Dredze, M., Ferryman, K., Gichoya, J.~W., Jurafsky, D., Koh, P.~W., Levy, K., Mullainathan, S., Obermeyer, Z., Suresh, H., and Vafa, K.
\newblock Using {Large} {Language} {Models} to {Promote} {Health} {Equity}.
\newblock \emph{NEJM AI}, 2\penalty0 (2):\penalty0 AIp2400889, January 2025.
\newblock \doi{10.1056/AIp2400889}.
\newblock URL \url{https://ai.nejm.org/doi/full/10.1056/AIp2400889}.
\newblock Publisher: Massachusetts Medical Society.

\bibitem[Pour(2023)]{pour2023police}
Pour, S.
\newblock Police use of facial recognition technology and racial bias--an assessment of criticisms of its current use.
\newblock \emph{American Journal of Artificial Intelligence}, 7\penalty0 (1):\penalty0 17--23, 2023.

\bibitem[Putnam(2011)]{putnam_reconsideration_2011}
Putnam, H.
\newblock A {Reconsideration} of {Deweyan} {Democracy} ({Reprint} from 1989).
\newblock In \emph{The pragmatism reader: from {Peirce} through the present}. Princeton University Press, Princeton, NJ Oxford, 2011.
\newblock ISBN 978-0-691-13705-6 978-0-691-13706-3.

\bibitem[Raji et~al.(2021)Raji, Denton, Bender, Hanna, and Paullada]{raji_ai_2021}
Raji, D., Denton, E., Bender, E.~M., Hanna, A., and Paullada, A.
\newblock {AI} and the {Everything} in the {Whole} {Wide} {World} {Benchmark}.
\newblock \emph{Proceedings of the Neural Information Processing Systems Track on Datasets and Benchmarks}, 1, December 2021.
\newblock URL \url{https://datasets-benchmarks-proceedings.neurips.cc/paper/2021/hash/084b6fbb10729ed4da8c3d3f5a3ae7c9-Abstract-round2.html}.

\bibitem[Raji et~al.(2020)Raji, Gebru, Mitchell, Buolamwini, Lee, and Denton]{raji_saving_2020}
Raji, I.~D., Gebru, T., Mitchell, M., Buolamwini, J., Lee, J., and Denton, E.
\newblock Saving {Face}: {Investigating} the {Ethical} {Concerns} of {Facial} {Recognition} {Auditing}.
\newblock In \emph{Proceedings of the {AAAI}/{ACM} {Conference} on {AI}, {Ethics}, and {Society}}, {AIES} '20, pp.\  145--151, New York, NY, USA, February 2020. Association for Computing Machinery.
\newblock ISBN 978-1-4503-7110-0.
\newblock \doi{10.1145/3375627.3375820}.
\newblock URL \url{https://doi.org/10.1145/3375627.3375820}.

\bibitem[Raji et~al.(2022)Raji, Kumar, Horowitz, and Selbst]{raji_fallacy_2022}
Raji, I.~D., Kumar, I.~E., Horowitz, A., and Selbst, A.
\newblock The {Fallacy} of {AI} {Functionality}.
\newblock In \emph{2022 {ACM} {Conference} on {Fairness}, {Accountability}, and {Transparency}}, {FAccT} '22, pp.\  959--972, New York, NY, USA, June 2022. Association for Computing Machinery.
\newblock ISBN 978-1-4503-9352-2.
\newblock \doi{10.1145/3531146.3533158}.
\newblock URL \url{https://dl.acm.org/doi/10.1145/3531146.3533158}.

\bibitem[Rastogi et~al.(2022)Rastogi, Stelmakh, Shen, Meila, Echenique, Chawla, and Shah]{rastogi_arxiv_2022}
Rastogi, C., Stelmakh, I., Shen, X., Meila, M., Echenique, F., Chawla, S., and Shah, N.~B.
\newblock To {ArXiv} or not to {ArXiv}: {A} {Study} {Quantifying} {Pros} and {Cons} of {Posting} {Preprints} {Online}, June 2022.
\newblock URL \url{http://arxiv.org/abs/2203.17259}.
\newblock arXiv:2203.17259 [cs].

\bibitem[Rombach et~al.(2022)Rombach, Blattmann, Lorenz, Esser, and Ommer]{rombach2022highresolutionimagesynthesislatent}
Rombach, R., Blattmann, A., Lorenz, D., Esser, P., and Ommer, B.
\newblock High-resolution image synthesis with latent diffusion models.
\newblock In \emph{Proceedings of the IEEE/CVF Conference on Computer Vision and Pattern Recognition (CVPR)}, pp.\  10684--10695, June 2022.

\bibitem[Rossbach(2023)]{rossbach_innocent_2023}
Rossbach, N.
\newblock Innocent until {Predicted} {Guilty}: {How} {Premature} {Predictive} {Policing} {Can} {Lead} to a {Self}-{Fulfilling} {Prophecy} of {Juvenile} {Delinquency} {Note}.
\newblock \emph{Florida Law Review}, 75\penalty0 (1):\penalty0 167--194, 2023.
\newblock URL \url{https://heinonline.org/HOL/P?h=hein.journals/uflr75&i=167}.

\bibitem[{SAE International}(2021)]{sae_international_j3016_202104_2021}
{SAE International}.
\newblock J3016\_202104: {Taxonomy} and {Definitions} for {Terms} {Related} to {Driving} {Automation} {Systems} for {On}-{Road} {Motor} {Vehicles}, April 2021.
\newblock URL \url{https://www.sae.org/standards/content/j3016_202104/}.

\bibitem[Salavati et~al.(2024)Salavati, Song, Diaz, Hale, Montenegro, Murai, and Dori-Hacohen]{salavati_reducing_2024}
Salavati, C., Song, S., Diaz, W.~S., Hale, S.~A., Montenegro, R.~E., Murai, F., and Dori-Hacohen, S.
\newblock Reducing {Biases} towards {Minoritized} {Populations} in {Medical} {Curricular} {Content} via {Artificial} {Intelligence} for {Fairer} {Health} {Outcomes}.
\newblock \emph{Proceedings of the AAAI/ACM Conference on AI, Ethics, and Society}, 7\penalty0 (1):\penalty0 1269--1280, October 2024.
\newblock ISSN 3065-8365.
\newblock \doi{10.1609/aies.v7i1.31722}.
\newblock URL \url{https://ojs.aaai.org/index.php/AIES/article/view/31722}.
\newblock Number: 1.

\bibitem[Salem et~al.(2024)Salem, Azzam, Emam, and Abohany]{salem_advancing_2024}
Salem, A.~H., Azzam, S.~M., Emam, O.~E., and Abohany, A.~A.
\newblock Advancing cybersecurity: a comprehensive review of {AI}-driven detection techniques.
\newblock \emph{Journal of Big Data}, 11\penalty0 (1):\penalty0 105, August 2024.
\newblock ISSN 2196-1115.
\newblock \doi{10.1186/s40537-024-00957-y}.
\newblock URL \url{https://doi.org/10.1186/s40537-024-00957-y}.

\bibitem[Salvaggio(2025)]{salvaggio_most_2025}
Salvaggio, E.
\newblock Most {Researchers} {Do} {Not} {Believe} {AGI} {Is} {Imminent}. {Why} {Do} {Policymakers} {Act} {Otherwise}? {\textbar} {TechPolicy}.{Press}, March 2025.
\newblock URL \url{https://techpolicy.press/most-researchers-do-not-believe-agi-is-imminent-why-do-policymakers-act-otherwise}.

\bibitem[Sartori \& Bocca(2023)Sartori and Bocca]{sartori_minding_2023}
Sartori, L. and Bocca, G.
\newblock Minding the gap(s): public perceptions of {AI} and socio-technical imaginaries.
\newblock \emph{AI \& SOCIETY}, 38\penalty0 (2):\penalty0 443--458, April 2023.
\newblock ISSN 1435-5655.
\newblock \doi{10.1007/s00146-022-01422-1}.
\newblock URL \url{https://doi.org/10.1007/s00146-022-01422-1}.

\bibitem[Saxon et~al.(2024)Saxon, Holtzman, West, Wang, and Saphra]{saxon2024benchmarksmicroscopesmodelmetrology}
Saxon, M., Holtzman, A., West, P., Wang, W.~Y., and Saphra, N.
\newblock Benchmarks as microscopes: A call for model metrology.
\newblock In \emph{First Conference on Language Modeling}, 2024.
\newblock URL \url{https://openreview.net/forum?id=bttKwCZDkm}.

\bibitem[Scheuerman et~al.(2021)Scheuerman, Hanna, and Denton]{scheuerman2021datasets}
Scheuerman, M.~K., Hanna, A., and Denton, E.
\newblock Do datasets have politics? {D}isciplinary values in computer vision dataset development.
\newblock \emph{Proceedings of the ACM on Human-Computer Interaction}, 5\penalty0 (CSCW):\penalty0 1--37, 2021.

\bibitem[Schulz et~al.(2002)Schulz, Krieger, and Galea]{schulz_addressing_2002}
Schulz, A.~J., Krieger, J., and Galea, S.
\newblock Addressing {Social} {Determinants} of {Health}: {Community}-{Based} {Participatory} {Approaches} to {Research} and {Practice}.
\newblock \emph{Health Education \& Behavior}, 29\penalty0 (3):\penalty0 287--295, June 2002.
\newblock ISSN 1090-1981.
\newblock \doi{10.1177/109019810202900302}.
\newblock URL \url{https://doi.org/10.1177/109019810202900302}.
\newblock Publisher: SAGE Publications Inc.

\bibitem[Sculley et~al.(2014)Sculley, Holt, Golovin, Davydov, Phillips, Ebner, Chaudhary, and Young]{sculley2014machine}
Sculley, D., Holt, G., Golovin, D., Davydov, E., Phillips, T., Ebner, D., Chaudhary, V., and Young, M.
\newblock Machine learning: The high interest credit card of technical debt.
\newblock In \emph{SE4ML: software engineering for machine learning (NIPS 2014 Workshop)}, volume 111, pp.\  112. Cambridge, MA, 2014.

\bibitem[Searle(1980)]{searle_minds_1980}
Searle, J.~R.
\newblock Minds, brains, and programs.
\newblock \emph{Behavioral and Brain Sciences}, 3\penalty0 (3):\penalty0 417--424, September 1980.
\newblock ISSN 1469-1825, 0140-525X.
\newblock \doi{10.1017/S0140525X00005756}.
\newblock URL \url{https://www.cambridge.org/core/journals/behavioral-and-brain-sciences/article/minds-brains-and-programs/DC644B47A4299C637C89772FACC2706A}.

\bibitem[Selbst et~al.(2019)Selbst, Boyd, Friedler, Venkatasubramanian, and Vertesi]{selbst_fairness_2019}
Selbst, A.~D., Boyd, D., Friedler, S.~A., Venkatasubramanian, S., and Vertesi, J.
\newblock Fairness and {Abstraction} in {Sociotechnical} {Systems}.
\newblock In \emph{Proceedings of the {Conference} on {Fairness}, {Accountability}, and {Transparency}}, pp.\  59--68, Atlanta GA USA, January 2019. ACM.
\newblock ISBN 978-1-4503-6125-5.
\newblock \doi{10.1145/3287560.3287598}.
\newblock URL \url{https://dl.acm.org/doi/10.1145/3287560.3287598}.

\bibitem[Sevilla et~al.(2022)Sevilla, Heim, Ho, Besiroglu, Hobbhahn, and Villalobos]{Sevilla_2022}
Sevilla, J., Heim, L., Ho, A., Besiroglu, T., Hobbhahn, M., and Villalobos, P.
\newblock Compute trends across three eras of machine learning.
\newblock In \emph{2022 International Joint Conference on Neural Networks (IJCNN)}, pp.\  1–8. IEEE, July 2022.
\newblock \doi{10.1109/ijcnn55064.2022.9891914}.
\newblock URL \url{http://dx.doi.org/10.1109/IJCNN55064.2022.9891914}.

\bibitem[Shelby et~al.(2023)Shelby, Rismani, Henne, Moon, Rostamzadeh, Nicholas, Yilla-Akbari, Gallegos, Smart, Garcia, and Virk]{shelby_sociotechnical_2023}
Shelby, R., Rismani, S., Henne, K., Moon, A., Rostamzadeh, N., Nicholas, P., Yilla-Akbari, N., Gallegos, J., Smart, A., Garcia, E., and Virk, G.
\newblock Sociotechnical {Harms} of {Algorithmic} {Systems}: {Scoping} a {Taxonomy} for {Harm} {Reduction}.
\newblock In \emph{Proceedings of the 2023 {AAAI}/{ACM} {Conference} on {AI}, {Ethics}, and {Society}}, {AIES} '23, pp.\  723--741, New York, NY, USA, August 2023. Association for Computing Machinery.
\newblock ISBN 9798400702310.
\newblock \doi{10.1145/3600211.3604673}.
\newblock URL \url{https://doi.org/10.1145/3600211.3604673}.

\bibitem[Shi \& Evans(2023)Shi and Evans]{shi_surprising_2023}
Shi, F. and Evans, J.
\newblock Surprising combinations of research contents and contexts are related to impact and emerge with scientific outsiders from distant disciplines.
\newblock \emph{Nature Communications}, 14\penalty0 (1):\penalty0 1641, March 2023.
\newblock ISSN 2041-1723.
\newblock \doi{10.1038/s41467-023-36741-4}.
\newblock URL \url{https://www.nature.com/articles/s41467-023-36741-4}.
\newblock Publisher: Nature Publishing Group.

\bibitem[Shilton(2018)]{shilton_values_2018}
Shilton, K.
\newblock Values and {Ethics} in {Human}-{Computer} {Interaction}.
\newblock \emph{Foundations and Trends® in Human–Computer Interaction}, 12\penalty0 (2):\penalty0 107--171, 2018.
\newblock ISSN 1551-3955, 1551-3963.
\newblock \doi{10.1561/1100000073}.
\newblock URL \url{http://www.nowpublishers.com/article/Details/HCI-073}.

\bibitem[Siler et~al.(2015)Siler, Lee, and Bero]{SilerEtAl2014}
Siler, K., Lee, K., and Bero, L.
\newblock Measuring the effectiveness of scientific gatekeeping.
\newblock \emph{Proceedings of the National Academy of Sciences}, 112\penalty0 (2):\penalty0 360--365, 2015.
\newblock \doi{10.1073/pnas.1418218112}.
\newblock URL \url{https://www.pnas.org/doi/abs/10.1073/pnas.1418218112}.

\bibitem[Simonton(2004)]{simonton_psychologys_2004}
Simonton, D.~K.
\newblock Psychology's {Status} as a {Scientific} {Discipline}: {Its} {Empirical} {Placement} within an {Implicit} {Hierarchy} of the {Sciences}.
\newblock \emph{Review of General Psychology}, 8\penalty0 (1):\penalty0 59--67, March 2004.
\newblock ISSN 1089-2680.
\newblock \doi{10.1037/1089-2680.8.1.59}.
\newblock URL \url{https://doi.org/10.1037/1089-2680.8.1.59}.

\bibitem[Sloane et~al.(2022)Sloane, Moss, and Chowdhury]{sloane_silicon_2022}
Sloane, M., Moss, E., and Chowdhury, R.
\newblock A {Silicon} {Valley} love triangle: {Hiring} algorithms, pseudo-science, and the quest for auditability.
\newblock \emph{Patterns}, 3\penalty0 (2), February 2022.
\newblock ISSN 2666-3899.
\newblock \doi{10.1016/j.patter.2021.100425}.
\newblock URL \url{https://www.cell.com/patterns/abstract/S2666-3899(21)00308-1}.
\newblock Publisher: Elsevier.

\bibitem[Smart(2015)]{smart_beyond_2015}
Smart, A.
\newblock \emph{Beyond zero and one: machines, psychedelics, and consciousness}.
\newblock OR Books, New York, 2015.
\newblock ISBN 978-1-68219-006-7.

\bibitem[Soderberg et~al.(2020)Soderberg, Errington, and Nosek]{soderberg_credibility_2020}
Soderberg, C.~K., Errington, T.~M., and Nosek, B.~A.
\newblock Credibility of preprints: an interdisciplinary survey of researchers.
\newblock \emph{Royal Society Open Science}, 7\penalty0 (10):\penalty0 201520, October 2020.
\newblock \doi{10.1098/rsos.201520}.
\newblock URL \url{https://royalsocietypublishing.org/doi/full/10.1098/rsos.201520}.
\newblock Publisher: Royal Society.

\bibitem[Sorensen et~al.(2024{\natexlab{a}})Sorensen, Jiang, Hwang, Levine, Pyatkin, West, Dziri, Lu, Rao, Bhagavatula, Sap, Tasioulas, and Choi]{sorensen_value_2024}
Sorensen, T., Jiang, L., Hwang, J.~D., Levine, S., Pyatkin, V., West, P., Dziri, N., Lu, X., Rao, K., Bhagavatula, C., Sap, M., Tasioulas, J., and Choi, Y.
\newblock Value {Kaleidoscope}: {Engaging} {AI} with {Pluralistic} {Human} {Values}, {Rights}, and {Duties}.
\newblock \emph{Proceedings of the AAAI Conference on Artificial Intelligence}, 38\penalty0 (18):\penalty0 19937--19947, March 2024{\natexlab{a}}.
\newblock ISSN 2374-3468.
\newblock \doi{10.1609/aaai.v38i18.29970}.
\newblock URL \url{https://ojs.aaai.org/index.php/AAAI/article/view/29970}.
\newblock Number: 18.

\bibitem[Sorensen et~al.(2024{\natexlab{b}})Sorensen, Moore, Fisher, Gordon, Mireshghallah, Rytting, Ye, Jiang, Lu, Dziri, Althoff, and Choi]{sorensen_roadmap_2024}
Sorensen, T., Moore, J., Fisher, J., Gordon, M.~L., Mireshghallah, N., Rytting, C.~M., Ye, A., Jiang, L., Lu, X., Dziri, N., Althoff, T., and Choi, Y.
\newblock Position: A roadmap to pluralistic alignment.
\newblock In \emph{Forty-first International Conference on Machine Learning}, 2024{\natexlab{b}}.
\newblock URL \url{https://openreview.net/forum?id=gQpBnRHwxM}.

\bibitem[Srivastava et~al.(2024)Srivastava, Chou, Shroff, Livescu, and Graziul]{srivastava_speech_2024}
Srivastava, T., Chou, J.-C., Shroff, P., Livescu, K., and Graziul, C.
\newblock Speech {Recognition} {For} {Analysis} of {Police} {Radio} {Communication}.
\newblock In \emph{2024 {IEEE} {Spoken} {Language} {Technology} {Workshop} ({SLT})}, pp.\  906--912, December 2024.
\newblock \doi{10.1109/SLT61566.2024.10832157}.
\newblock URL \url{https://ieeexplore.ieee.org/document/10832157/metrics#metrics}.

\bibitem[Stirling(2014)]{stirling2014disciplinary}
Stirling, A.
\newblock Disciplinary dilemma: working across research silos is harder than it looks.
\newblock \emph{The Guardian}, 11:\penalty0 1--4, 2014.

\bibitem[Stokols et~al.(2003)Stokols, Fuqua, Gress, Harvey, Phillips, Baezconde-Garbanati, Unger, Palmer, Clark, Colby, et~al.]{stokols2003evaluating}
Stokols, D., Fuqua, J., Gress, J., Harvey, R., Phillips, K., Baezconde-Garbanati, L., Unger, J., Palmer, P., Clark, M.~A., Colby, S.~M., et~al.
\newblock Evaluating transdisciplinary science.
\newblock \emph{Nicotine \& tobacco research}, 5\penalty0 (Suppl\_1):\penalty0 S21--S39, 2003.

\bibitem[Suchman(2023)]{suchman_uncontroversial_2023}
Suchman, L.
\newblock The uncontroversial ‘thingness’ of {AI}.
\newblock \emph{Big Data \& Society}, 10\penalty0 (2):\penalty0 20539517231206794, July 2023.
\newblock ISSN 2053-9517.
\newblock \doi{10.1177/20539517231206794}.
\newblock URL \url{https://doi.org/10.1177/20539517231206794}.
\newblock Publisher: SAGE Publications Ltd.

\bibitem[Suleyman \& Bhaskar(2023)Suleyman and Bhaskar]{suleyman_coming_2023}
Suleyman, M. and Bhaskar, M.
\newblock \emph{The {Coming} {Wave}}.
\newblock Crown, New York, first edition edition, 2023.
\newblock ISBN 978-0-593-59396-7.

\bibitem[Summerfield(2023)]{summerfield_natural_2023}
Summerfield, C.
\newblock \emph{Natural general intelligence: how understanding the brain can help us build {AI}}.
\newblock Oxford University Press, Oxford New York, NY, first edition edition, 2023.
\newblock ISBN 978-0-19-284388-3.

\bibitem[Tenopir et~al.(2016)Tenopir, Levine, Allard, Christian, Volentine, Boehm, Nichols, Nicholas, Jamali, Herman, and Watkinson]{tenopir_trustworthiness_2016}
Tenopir, C., Levine, K., Allard, S., Christian, L., Volentine, R., Boehm, R., Nichols, F., Nicholas, D., Jamali, H.~R., Herman, E., and Watkinson, A.
\newblock Trustworthiness and authority of scholarly information in a digital age: {Results} of an international questionnaire.
\newblock \emph{Journal of the Association for Information Science and Technology}, 67\penalty0 (10):\penalty0 2344--2361, 2016.
\newblock ISSN 2330-1643.
\newblock \doi{10.1002/asi.23598}.
\newblock URL \url{https://onlinelibrary.wiley.com/doi/abs/10.1002/asi.23598}.

\bibitem[{The Royal Society}(2024)]{royal2024science}
{The Royal Society}.
\newblock \emph{Science in the age of AI: How artificial intelligence is changing the nature and method of scientific research}.
\newblock The Royal Society, United Kingdom, May 2024.

\bibitem[Tibebu(2025)]{tibebu_deepseek_2025}
Tibebu, H.
\newblock {DeepSeek} and the {Race} to {AGI}: {How} {Global} {AI} {Competition} {Puts} {Ethical} {Accountability} at {Risk} {\textbar} {TechPolicy}.{Press}.
\newblock \emph{Tech Policy Press}, January 2025.
\newblock URL \url{https://techpolicy.press/deepseek-and-the-race-to-agi-how-global-ai-competition-puts-ethical-accountability-at-risk}.

\bibitem[{United Nations}(2024)]{united_nations_governing_2024}
{United Nations}.
\newblock Governing {AI} for {Humanity}.
\newblock Final {Report}, United Nations, New York, NY, September 2024.
\newblock URL \url{https://www.un.org/sites/un2.un.org/files/governing_ai_for_humanity_final_report_en.pdf}.

\bibitem[Van~Rooij et~al.(2024)Van~Rooij, Guest, Adolfi, de~Haan, Kolokolova, and Rich]{van2024reclaiming}
Van~Rooij, I., Guest, O., Adolfi, F., de~Haan, R., Kolokolova, A., and Rich, P.
\newblock Reclaiming {AI} as a theoretical tool for cognitive science.
\newblock \emph{Computational Brain \& Behavior}, pp.\  1--21, 2024.

\bibitem[Veit(2020)]{veit_model_2020}
Veit, W.
\newblock Model {Pluralism}.
\newblock \emph{Philosophy of the Social Sciences}, 50\penalty0 (2):\penalty0 91--114, March 2020.
\newblock ISSN 0048-3931, 1552-7441.
\newblock \doi{10.1177/0048393119894897}.
\newblock URL \url{http://journals.sagepub.com/doi/10.1177/0048393119894897}.

\bibitem[Venkit et~al.(2024)Venkit, Graziul, Goodman, Kenny, and Wilson]{venkit_race_2024}
Venkit, P.~N., Graziul, C., Goodman, M.~A., Kenny, S.~N., and Wilson, S.
\newblock Race and {Privacy} in {Broadcast} {Police} {Communications}.
\newblock \emph{Proc. ACM Hum.-Comput. Interact.}, 8\penalty0 (CSCW2):\penalty0 382:1--382:26, November 2024.
\newblock \doi{10.1145/3686921}.
\newblock URL \url{https://dl.acm.org/doi/10.1145/3686921}.

\bibitem[Vestal \& Mesmer-Magnus(2020)Vestal and Mesmer-Magnus]{vestal2020interdisciplinarity}
Vestal, A. and Mesmer-Magnus, J.
\newblock Interdisciplinarity and team innovation: The role of team experiential and relational resources.
\newblock \emph{Small Group Research}, 51\penalty0 (6):\penalty0 738--775, 2020.

\bibitem[Viljoen(2021)]{viljoen_relational_2021}
Viljoen, S.
\newblock A {Relational} {Theory} of {Data} {Governance}.
\newblock \emph{The Yale Law Journal}, 2021.

\bibitem[Wang et~al.(2024)Wang, Kapoor, Barocas, and Narayanan]{wang_against_2024}
Wang, A., Kapoor, S., Barocas, S., and Narayanan, A.
\newblock Against {Predictive} {Optimization}: {On} the {Legitimacy} of {Decision}-making {Algorithms} {That} {Optimize} {Predictive} {Accuracy}.
\newblock \emph{ACM Journal on Responsible Computing}, 1\penalty0 (1):\penalty0 1--45, March 2024.
\newblock ISSN 2832-0565.
\newblock \doi{10.1145/3636509}.
\newblock URL \url{https://dl.acm.org/doi/10.1145/3636509}.

\bibitem[Wang et~al.(2022)Wang, Prabhat, and Sambasivan]{wang_whose_2022}
Wang, D., Prabhat, S., and Sambasivan, N.
\newblock Whose {AI} {Dream}? {In} search of the aspiration in data annotation.
\newblock In \emph{Proceedings of the 2022 {CHI} {Conference} on {Human} {Factors} in {Computing} {Systems}}, {CHI} '22, pp.\  1--16, New York, NY, USA, April 2022. Association for Computing Machinery.
\newblock ISBN 978-1-4503-9157-3.
\newblock \doi{10.1145/3491102.3502121}.
\newblock URL \url{https://dl.acm.org/doi/10.1145/3491102.3502121}.

\bibitem[Warne \& Burningham(2019)Warne and Burningham]{warne_spearmans_2019}
Warne, R.~T. and Burningham, C.
\newblock Spearman’s g found in 31 non-{Western} nations: {Strong} evidence that g is a universal phenomenon.
\newblock \emph{Psychological Bulletin}, 145\penalty0 (3):\penalty0 237--272, March 2019.
\newblock ISSN 0033-2909.
\newblock \doi{10.1037/bul0000184}.
\newblock URL \url{http://proxy.uchicago.edu/login?url=https://search.ebscohost.com/login.aspx?direct=true&db=pdh&AN=2019-01683-001&site=ehost-live&scope=site}.

\bibitem[Weidinger et~al.(2024)Weidinger, Mellor, Pegueroles, Marchal, Kumar, Lum, Akbulut, Diaz, Bergman, Rodriguez, Rieser, and Isaac]{weidinger_star_2024}
Weidinger, L., Mellor, J. F.~J., Pegueroles, B.~G., Marchal, N., Kumar, R., Lum, K., Akbulut, C., Diaz, M., Bergman, A.~S., Rodriguez, M.~D., Rieser, V., and Isaac, W.
\newblock {STAR}: {SocioTechnical} {Approach} to {Red} {Teaming} {Language} {Models}.
\newblock In Al-Onaizan, Y., Bansal, M., and Chen, Y.-N. (eds.), \emph{Proceedings of the 2024 {Conference} on {Empirical} {Methods} in {Natural} {Language} {Processing}}, pp.\  21516--21532, Miami, Florida, USA, November 2024. Association for Computational Linguistics.
\newblock \doi{10.18653/v1/2024.emnlp-main.1200}.
\newblock URL \url{https://aclanthology.org/2024.emnlp-main.1200/}.

\bibitem[Weizenbaum(1976)]{weizenbaum_computer_1976}
Weizenbaum, J.
\newblock \emph{Computer power and human reason: from judgment to calculation}.
\newblock Freeman, San Francisco, 1976.
\newblock ISBN 978-0-7167-0464-5 978-0-7167-0463-8.

\bibitem[Whitney \& Norman(2024)Whitney and Norman]{whitney_real_2024}
Whitney, C.~D. and Norman, J.
\newblock Real {Risks} of {Fake} {Data}: {Synthetic} {Data}, {Diversity}-{Washing} and {Consent} {Circumvention}.
\newblock In \emph{Proceedings of the 2024 {ACM} {Conference} on {Fairness}, {Accountability}, and {Transparency}}, {FAccT} '24, pp.\  1733--1744, New York, NY, USA, June 2024. Association for Computing Machinery.
\newblock ISBN 9798400704505.
\newblock \doi{10.1145/3630106.3659002}.
\newblock URL \url{https://dl.acm.org/doi/10.1145/3630106.3659002}.

\bibitem[Widder(2024)]{widder_epistemic_2024}
Widder, D.~G.
\newblock Epistemic {Power} in {AI} {Ethics} {Labor}: {Legitimizing} {Located} {Complaints}.
\newblock In \emph{Proceedings of the 2024 {ACM} {Conference} on {Fairness}, {Accountability}, and {Transparency}}, {FAccT} '24, pp.\  1295--1304, New York, NY, USA, June 2024. Association for Computing Machinery.
\newblock ISBN 9798400704505.
\newblock \doi{10.1145/3630106.3658973}.
\newblock URL \url{https://dl.acm.org/doi/10.1145/3630106.3658973}.

\bibitem[Widder \& Hicks(2024)Widder and Hicks]{widder_watching_2024}
Widder, D.~G. and Hicks, M.
\newblock Watching the {Generative} {AI} {Hype} {Bubble} {Deflate}, August 2024.
\newblock URL \url{http://arxiv.org/abs/2408.08778}.
\newblock arXiv:2408.08778.

\bibitem[Widder \& Nafus(2023)Widder and Nafus]{widder_dislocated_2023}
Widder, D.~G. and Nafus, D.
\newblock Dislocated accountabilities in the “{AI} supply chain”: {Modularity} and developers’ notions of responsibility.
\newblock \emph{Big Data \& Society}, 10\penalty0 (1):\penalty0 20539517231177620, January 2023.
\newblock ISSN 2053-9517.
\newblock \doi{10.1177/20539517231177620}.
\newblock URL \url{https://doi.org/10.1177/20539517231177620}.
\newblock Publisher: SAGE Publications Ltd.

\bibitem[Xu et~al.(2022)Xu, Wu, and Evans]{xu_flat_2022}
Xu, F., Wu, L., and Evans, J.
\newblock Flat teams drive scientific innovation.
\newblock \emph{Proceedings of the National Academy of Sciences}, 119\penalty0 (23):\penalty0 e2200927119, June 2022.
\newblock \doi{10.1073/pnas.2200927119}.
\newblock URL \url{https://www.pnas.org/doi/abs/10.1073/pnas.2200927119}.
\newblock Publisher: Proceedings of the National Academy of Sciences.

\bibitem[Xu et~al.(2024)Xu, Wang, Fan, and Liu]{xu2024benchmarkingbenchmarkleakagelarge}
Xu, R., Wang, Z., Fan, R.-Z., and Liu, P.
\newblock Benchmarking benchmark leakage in large language models, 2024.
\newblock URL \url{https://arxiv.org/abs/2404.18824}.

\bibitem[Young et~al.(2019)Young, Rodriguez, Keller, Sun, Sa, Whittington, and Howe]{young_beyond_2019}
Young, M., Rodriguez, L., Keller, E., Sun, F., Sa, B., Whittington, J., and Howe, B.
\newblock Beyond {Open} vs. {Closed}: {Balancing} {Individual} {Privacy} and {Public} {Accountability} in {Data} {Sharing}.
\newblock In \emph{Proceedings of the {Conference} on {Fairness}, {Accountability}, and {Transparency}}, {FAT}* '19, pp.\  191--200, New York, NY, USA, January 2019. Association for Computing Machinery.
\newblock ISBN 978-1-4503-6125-5.
\newblock \doi{10.1145/3287560.3287577}.
\newblock URL \url{https://dl.acm.org/doi/10.1145/3287560.3287577}.

\bibitem[Young et~al.(2024)Young, Ehsan, Singh, Tafesse, Gilman, Harrington, and Metcalf]{young_participation_2024}
Young, M., Ehsan, U., Singh, R., Tafesse, E., Gilman, M., Harrington, C., and Metcalf, J.
\newblock Participation versus scale: {Tensions} in the practical demands on participatory {AI}.
\newblock \emph{First Monday}, April 2024.
\newblock ISSN 1396-0466.
\newblock \doi{20240428092301000}.
\newblock URL \url{https://firstmonday.org/ojs/index.php/fm/article/view/13642}.

\bibitem[Yu et~al.(2023)Yu, Rosenfeld, and Gupta]{yu_ai_2023}
Yu, D., Rosenfeld, H., and Gupta, A.
\newblock The ‘{AI} divide’ between the {Global} {North} and {Global} {South}, January 2023.
\newblock URL \url{https://www.weforum.org/stories/2023/01/davos23-ai-divide-global-north-global-south/}.

\bibitem[Zeff(2025)]{zeff2024microsoft}
Zeff, M.
\newblock Microsoft and {OpenAI} have a financial definition of {AGI}: Report, 2025.
\newblock URL \url{https://techcrunch.com/2024/12/26/microsoft-and-openai-have-a-financial-definition-of-agi-report/}.
\newblock [Online; accessed 17-January-2025].

\bibitem[Zhang et~al.(2024)Zhang, Da, Lee, Robinson, Wu, Song, Zhao, Raja, Zhuang, Slack, Lyu, Hendryx, Kaplan, Lunati, and Yue]{zhang2024careful}
Zhang, H., Da, J., Lee, D., Robinson, V., Wu, C., Song, W., Zhao, T., Raja, P., Zhuang, C., Slack, D., Lyu, Q., Hendryx, S., Kaplan, R., Lunati, M., and Yue, S.
\newblock A {Careful} {Examination} of {Large} {Language} {Model} {Performance} on {Grade} {School} {Arithmetic}.
\newblock \emph{Advances in Neural Information Processing Systems}, 37:\penalty0 46819--46836, December 2024.
\newblock URL \url{https://proceedings.neurips.cc/paper_files/paper/2024/hash/53384f2090c6a5cac952c598fd67992f-Abstract-Datasets_and_Benchmarks_Track.html}.

\bibitem[Zhang et~al.(2021)Zhang, Sun, Jiang, and Huang]{zhang_relationship_2021}
Zhang, L., Sun, B., Jiang, L., and Huang, Y.
\newblock On the relationship between interdisciplinarity and impact: {Distinct} effects on academic and broader impact.
\newblock \emph{Research Evaluation}, 30\penalty0 (3):\penalty0 256--268, July 2021.
\newblock ISSN 0958-2029.
\newblock \doi{10.1093/reseval/rvab007}.
\newblock URL \url{https://doi.org/10.1093/reseval/rvab007}.

\bibitem[Zhao et~al.(2024)Zhao, Andrews, Papakyriakopoulos, and Xiang]{pmlr-v235-zhao24a}
Zhao, D., Andrews, J., Papakyriakopoulos, O., and Xiang, A.
\newblock Position: Measure dataset diversity, don’t just claim it.
\newblock In Salakhutdinov, R., Kolter, Z., Heller, K., Weller, A., Oliver, N., Scarlett, J., and Berkenkamp, F. (eds.), \emph{Proceedings of the 41st International Conference on Machine Learning}, volume 235 of \emph{Proceedings of Machine Learning Research}, pp.\  60644--60673. PMLR, 21--27 Jul 2024.
\newblock URL \url{https://proceedings.mlr.press/v235/zhao24a.html}.

\bibitem[Zhu(2022)]{zhu_paradigm_2022}
Zhu, Z.
\newblock Paradigm, specialty, pragmatism: {Kuhn}'s legacy to methodological pluralism.
\newblock \emph{Systems Research and Behavioral Science}, 39\penalty0 (5):\penalty0 895--912, 2022.
\newblock ISSN 1099-1743.
\newblock \doi{10.1002/sres.2881}.
\newblock URL \url{https://onlinelibrary.wiley.com/doi/abs/10.1002/sres.2881}.
\newblock \_eprint: https://onlinelibrary.wiley.com/doi/pdf/10.1002/sres.2881.

\end{thebibliography}

\newpage
\appendix

\section{Definitions of AGI and Related Concepts}
\label{sec:definition}

\textbf{Table 1} below presents illustrative definitions of AGI and usefully related concepts. We agree with \citet{morris_position_2024}'s proposal to broaden discussions of AGI definitions to include accounts that avoid the term ``AGI'' yet address similar goals of achieving human-level intelligence. For example, while OpenAI's influential definition \cite{openai_2018} focuses on outperforming humans at economically valuable work, sharing key parallels with \citet{nilsson_human-level_2005}. Yet it notably differs from \citet{chollet_arc_technical_report_2024, summerfield_natural_2023, morris_position_2024, chollet_measure_2019, goertzel_artificial_2014} and others by not explicitly emphasizing generality. 

Following \citet{blili-hamelin_unsocial_2024}, we believe discussions of AGI definition should include approaches that challenge AGI's central premises. Below, we include \citet{weizenbaum_computer_1976} and \citet{attard-frost_queering_2023}. Including these critical accounts enables noticing a surprising similarity with \citet{summerfield_natural_2023}'s reconceptualization of AGI through the lens of natural intelligence: all three accounts favor a strong form of contextualism and pluralism about what intelligence means.

\section{AGI as a North-Star Goal}
\label{sec:north-star}

This paper argues against AGI serving as a north-star goal of AI research. When we talk about AGI being treated as north-star goal, we are not claiming that the majority of AI researchers are explicitly working in pursuit of this goal. In fact, many AI researchers may, like us, doubt or reject this goal \cite{salvaggio_most_2025,francesca_rossi_aaai_2025}. Rather, we claim that influential researchers and executives hold this view, enough so for it to deserve scrutiny. These dominant voices are further amplified by the publicity that discussions of AGI generate, including by members of the press \cite{klein_opinion_2025}, and by government commissions \cite{rishi_bommasani_draft_2025}. As such, AGI has come to permeate both community incentives and cultural norms. In this context, interrogating its role and influence in the AI research community matters. Below we provide a small sample of quotes and resources illustrating this effect.

\paragraph{OpenAI} The mission statement of OpenAI is ``\textellipsis to ensure that [AGI] benefits all of humanity.'' Its website~\cite{OpenAIAbout} states that ``we are building safe and beneficial AGI, but will also consider our mission fulfilled if our work aids others to achieve this outcome.'' This goal directly influences both the company's direct work~\cite{OpenAIPlanning} and the work that it funds~\cite{OpenAISecurity}.

\paragraph{Google DeepMind} The vision statement~\cite{DeepMindAbout2025} of Google DeepMind states that ``[AGI] has the potential to drive one of the greatest transformations in history.'' In a recent briefing~\cite{CNBC-AGI}, Demis Hassabis stated that, though current systems still have limitations, over the next 5--10 years ``a lot of those capabilities will start coming to the fore and we'll start moving towards what we call [AGI].'' A position paper~\cite{morris_position_2024} by Google DeepMind authors last year defines concrete goals in pursuit of AGI.

\paragraph{Anthropic}
In the essay ``Machines of Loving Grace,'' Anthropic CEO Dario Amodei argues how the world could be shaped positively by ``Powerful AI'' aligned with ``AGI'' goals~~\cite{machines-grace}. Amodei recently told CNBC that AI that is ``better than almost all humans at almost all tasks'' can emerge shortly~\cite{CNBC-AGI}.
Anthropic's official recommendations~\cite{anthropic-ostp} to OSTP for the U.S. AI Action Plan state that ``we expect powerful AI systems will emerge [with] intellectual capabilities matching or exceeding that of Nobel Prize winners.''

\paragraph{Other Influential Executives \& Researchers}
In ``Sparks of AGI''~\cite{bubeck2023sparksartificialgeneralintelligence}, researchers at Microsoft argue that GPT-4 ``could reasonably be viewed as an early (yet still incomplete) version of [AGI].''
In announcing xAI, Elon Musk stated~\cite{musk-xai} that ``the overarching goal of xAI is to build a good AGI.'' Speaking to TIME~\cite{time-meta}, Yann LeCunn explained that he refers to ``what people call `AGI''' as ``human-level intelligence,'' and noted that ``the mission of FAIR [Meta’s Fundamental AI Research team] is human-level intelligence.'' Geoff Hinton 
stated to Forbes~\cite{forbes-hinton} that he ``is certain we will have AGI soon, and biological humans will be relegated to be the second-smartest species on the planet.''

\onecolumn
\begin{center}
    \small
    \begin{longtable}{p{0.99\linewidth}}
        \caption{Sample of proposed definitions of AGI and related concepts.}
    \label{tab:agi_definitions} \\
    \toprule
    
\citet{weizenbaum_computer_1976}. ``Intelligence is a meaningless concept in and of itself. It requires a frame of reference, a specification of a domain of thought and action, in order to make it meaningful. [\textellipsis] [T]hese domains are themselves not measurable.'' Argues that any argument that calls for the conclusion or denial that ``machines may surpass us in general intelligence'' is ``ill-framed and therefore sterile'' due to ``our inability to compute an upper bound on machine intelligence.'' We follow \citet{blili-hamelin_unsocial_2024} in considering this critical account relevant to debates about how to conceive AGI. \\ \midrule

\citet{searle_minds_1980}. ``{according to strong AI, the computer is not merely a tool in the study of the mind; rather, the appropriately programmed computer really is a mind, in the sense that computers given the right programs can be literally said to understand and have other cognitive states}.'' \\ \midrule 

\citet{gubrud_nanotechnology_1997} ``By advanced artificial general intelligence, I mean AI systems that rival or surpass the human brain in complexity and speed, that can acquire, manipulate and reason with general knowledge, and that are usable in essentially any phase of industrial or military operations where a human intelligence would otherwise be needed. Such systems may be modeled on the human brain, but they do not necessarily have to be, and they do not have to be ``conscious'' or possess any other competence that is not strictly relevant to their application. What matters is that such systems can be used to replace human brains in tasks ranging from organizing and running a mine or a factory to piloting an airplane, analyzing intelligence data or planning a battle.'' \\ \midrule

\citet{nilsson_human-level_2005}. ``{achieving real human-level artificial intelligence would necessarily imply that most of the tasks that humans perform for pay could be automated. Rather than work toward this goal of automation by building special-purpose systems, I argue for the development of general-purpose, educable systems that can learn and be taught to perform any of the thousands of jobs that humans can perform.}'' \\ \midrule 

\citet{Wozniak_coffee_2010} ``\textit{Wozniak: Could a Computer Make a Cup of Coffee?}'' tasks the machine to go into an “average” American home, find ingredients, and make a cup of coffee. This requires embodied AI systems. Wozniak's test has since been included in discussions of AGI \cite{goertzel2012architecture}. \\ \midrule 

\citet{chalmers_singularity_2010}. ``AI is artificial intelligence of human level or greater (that is, at least as intelligent as an average human). Let us say that AI+ is artificial intelligence of greater than human level (that is, more intelligent than the most intelligent human). Let us say that AI++ (or superintelligence) is AI of far greater than human level (say, at least as far beyond the most intelligent human as the most intelligent human is beyond a mouse).'' \\ \midrule

\citet{goertzel2012architecture}. Propose an architecture for human-like general intelligence that integrates slightly modified versions of previously existing architectures, emphasizing the commonalities across different approaches.
\\ \midrule 

\citet{bostrom_superintelligence_2014}. ``We can tentatively define a superintelligence as \textit{any intellect that greatly exceeds the cognitive performance of humans in virtually all domains of interest}.'' 
\\ \midrule 

\citet{goertzel_artificial_2014}. ``roughly speaking, an AGI system is a synthetic intelligence that has a general scope and is good at generalization across various goals and contexts.'' 
\\ \midrule

\citet{smart_beyond_2015}. ``a strong AI system would be an entirely autonomous computer system in no way controlled or influenced by human operators. It could successfully adapt to its environment or even be part of its environment, making intelligent decisions, and for all intents and purposes interacting with humans naturally. It would have vastly superior memory and computational abilities but would also be able to reason and act accordingly. What all of this boils down to is that a strong AI would have to be conscious.'' \\ \midrule

\citet{openai_2018}. ``{OpenAI’s mission is to ensure that artificial general intelligence (AGI)—by which we mean highly autonomous systems that outperform humans at most economically valuable work—benefits all of humanity.}'' December 2024 reporting suggests that OpenAI and Microsoft ``signed an agreement last year stating OpenAI has only achieved AGI when it develops AI systems that can generate at least \$100 billion in profits'' \cite{zeff2024microsoft}. If true, this is a significant departure from their former definition. \\ \midrule 

\citet{chollet_measure_2019,chollet_arc_technical_report_2024}. Defines AGI as ``{a system capable of efficiently acquiring new skills and solving novel problems for which it was neither explicitly designed nor trained}.'' In 2019, introduced an as yet (January 2025) unsolved benchmark for incentivizing progress towards AGI thus defined. Proposes that ``it's still feasible to create unsaturated, interesting benchmarks that are easy for humans, yet impossible for AI -- without involving specialist knowledge. We will have AGI when creating such evals becomes outright impossible'' \cite{chollet_so_2024}. \\ \midrule 

\citet{hernandez-orallo_general_2021}. ``independently of its overall capability, \textit{an agent can only be called fully general if it covers all tasks up to an equivalent level of difficulty, determined by the resources that are needed for them}.'' Introduces two individual-specific (be them humans, other animals or AI systems) measures that, together, decouple the concept of general intelligence: (i) `generality', which refers to ``the \textit{distribution} of the tasks the agent can solve,'' and (ii) `capability', which indicates ``how far, on average, an agent can reach in terms of task difficulty.'' \\ \midrule 

\citet{Marcus_2022}. Defines AGI as {``a shorthand for any intelligence\textellipsis~that is flexible and general, with resourcefulness and reliability comparable to (or beyond) human intelligence.''} Calls for the need to operationalize the definition as a single system that can succeed in at least 3 of 5 proposed tasks, including Wozniak's coffee cup benchmark~\cite{Wozniak_coffee_2010}. \\ \midrule 

\citet{morris_position_2024}. Proposes a practical strategy analogous to Levels of Driving Automation standards \cite{sae_international_j3016_202104_2021}. Describes a graded set of levels of achievement of target characteristics that can each be associated with tangible ``metrics'', the introduction of ``risks'', and changes in ``Human-AI Interaction paradigm'' \cite{morris_position_2024}. The framework targets 2 characteristics: levels of {performance} (which they define as ``the depth of an AI system’s capabilities, i.e., how it compares to human-level performance for a given task''), and levels of {generality} (defined as ``breadth of an AI system’s capabilities, i.e., the range of tasks for which an AI system reaches a target performance threshold'').  \\ \midrule  

\citet{summerfield_natural_2023} We consider this account to endorse a strong form of {pluralism} about intelligence: natural intelligence takes {many} different shapes across cultures and species, serving {many} goals and functions that are irreducibly shaped by ``the internal model by which an animal understands the world'', which itself ``depends on its local environment, its embodied form, its desires and goals, and its interactions with conspecifics.'' The same should be expected for ``strong AI'' or ``AGI''. However, building on \citet{dreyfus_mind_1986}, Summerfield argues we have practical reasons to constrain the forms AI takes. The goal of AI is ``to help humans in their endeavours.'' To that end,``if we want to build AI systems that exhibit human-like intelligence, with whom we can interact in pursuit of human-centred goals, these agents will need to think in ways that make sense to us.''\footnote{``[W]e are building AI to make the world a better place. But if we want AI to be useful to people, it will need to share our umwelt. If we build an AI that sees the world in a radically different way to us, its behaviour and mental states will be unintelligible. Such an agent will be at best unreliable and at worst unsafe.'' \cite{summerfield_natural_2023}}  \\ \midrule

\citet{attard-frost_queering_2023} Defines human and artificial intelligence as ``value-dependent cognitive performance'', and ``centres interdependencies between agents, their environments, and their measurers in collectively constructing and measuring context-specific performances of intelligent action.'' Although this account is not presented as a conception of AGI, \citet{blili-hamelin_unsocial_2024} argue that the account is relevant to the topic.  \\ \midrule

\citet{suleyman_coming_2023} ``{Artificial intelligence (AI) is the science of teaching machines to learn humanlike capabilities. Artificial general intelligence (AGI) is the point at which an AI can perform all human cognitive skills better than the smartest humans.}'' \\ \midrule  

\citet{aguera_y_arcas_artificial_2023}. ``{`General intelligence' must be thought of in terms of a multidimensional scorecard, not a single yes/no proposition.}'' Dimensions discussed include topics, tasks, modalities, languages, and instructability. \\  
    \bottomrule \\
    \end{longtable}
\end{center}

\onecolumn

\section*{Author Contributions}

We follow the CRediT recommendations and taxonomy provided by
\citet{allen-etal-2019}
to determine and outline author contributions.\footnote{The initial ICML submission was created by Borhane Blili-Hamelin, Leif Hancox-Li, and Christopher Graziul, leveraging extensive work and writing from the larger project. All contributors then worked together on refining this submission.}
\begin{itemize}
    \item Borhane Blili-Hamelin: Conceptualization (Formulation \&~Evolution), Investigation, Methodology (Development), Project administration, Supervision (Oversight \&~Leadership), Writing (Initial draft, Submitted draft, Review \&~Editing).
    \item Christopher Graziul: Conceptualization (Formulation \&~Evolution), Investigation, Methodology (Development), Project administration, Supervision (Oversight \&~Leadership), Writing (Initial draft, Submitted draft, Review \&~Editing).
    \item Leif Hancox-Li: Conceptualization (Formulation \&~Ideas), Investigation, Methodology (Development), Supervision (Mentorship), Writing (Initial draft, Submitted draft, Review \&~Editing).
    \item Hananel Hazan: Conceptualization (Ideas \&~ Evolution), Investigation, Methodology (Development), Project administration, Writing (Initial draft, Review \&~Editing, \LaTeX).
    \item El-Mahdi El-Mhamdi: Conceptualization (Evolution \&~Ideas), Writing (Initial draft, Review \&~Editing).
    \item Avijit Ghosh: Conceptualization (Ideas), Writing (Submitted draft [Traps section], Review \&~Editing).
    \item Katherine Heller: Conceptualization (Formulation \&~Evolution), Supervision (Mentorship), Writing (Initial draft, Submitted draft, Review \&~Editing).
    \item Jacob Metcalf: Conceptualization (Formulation \&~Evolution), Writing (Submitted draft, Review \&~Editing).
    \item Fabricio Murai: Conceptualization (Ideas), Writing (Submitted draft [Table of AGI definitions], Review \&~Editing, \LaTeX).
    \item Eryk Salvaggio: Conceptualization (Evolution \&~Ideas), Writing (Submitted draft [Introduction, Traps section], Review \&~Editing).
    \item Andrew Smart: Conceptualization (Formulation \&~Evolution), Writing (Initial draft, Review \&~Editing).
    \item Todd Snider: Conceptualization (Ideas), Methodology (Implementation), Writing (Initial draft, Review \&~Editing, \LaTeX).
    \item Mariame Tighanimine: Conceptualization (Evolution \&~Ideas), Writing (Initial draft, Review \&~Editing).
    \item Talia Ringer: Conceptualization (Formulation, \&~Ideas), Project administration, Supervision (Oversight \&~Leadership), Writing (Initial draft, Review \&~Editing).
    \item Margaret Mitchell: Conceptualization (Ideas \&~ Evolution), Investigation, Methodology (Development), Supervision (Oversight \&~Mentorship), Writing (Initial draft, Submitted draft [all sections], Review \&~Editing, Rebuttal).
    \item Shiri Dori-Hacohen: Conceptualization (Ideas \&~Evolution), Investigation, Methodology, Project administration, Supervision (Oversight \&~Leadership), Writing (Initial draft, Submitted draft [all sections], Review \&~Editing).
\end{itemize}

\end{document}